\documentclass[10pt,journal,compsoc]{IEEEtran}
\usepackage{microtype}
\usepackage{graphicx}
\usepackage{subfigure}
\usepackage{booktabs} % for professional tables
\usepackage[frozencache,cachedir=.]{minted}
\usepackage{caption}
\usepackage{epstopdf}

\usepackage{amsmath,amssymb,amsfonts}
\usepackage{adjustbox}
\usepackage{graphicx}
\usepackage{textcomp}
\usepackage{comment}

\usepackage{hyperref}

\usepackage{multirow}% http://ctan.org/pkg/multirow
\usepackage{hhline}
\newlength\bibitemsep
\usepackage{listings}
\newcommand{\ignore}[1]{}

\begin{document}

%%%%%%%%%%%%%%%%%%
\newcommand{\name}{DeepFlow }
\title{DeepFlow: A Cross-Stack Pathfinding Framework for Distributed AI Systems}
%%%%%%%%%%%%%%%%%%
%\twocolumn[
%\mlsystitle{\papertitle}

%\and
%\and
%\and
%\and
%\and

%\maketitle
%\usepackage{savetrees}[moderate]

\author{Newsha~Ardalani,~\IEEEmembership{Member,~IEEE,}
        Saptadeep~Pal,~\IEEEmembership{Member,~IEEE,}
        and~Puneet~Gupta,~\IEEEmembership{Fellow,~IEEE}% <-this % stops a space
\IEEEcompsocitemizethanks{\IEEEcompsocthanksitem Newsha Ardalani is with Meta, Inc. This work was primarily done during her tenure at Baidu Research.\protect
E-mail: new@fb.com
\IEEEcompsocthanksitem Saptadeep Pal and Puneet Gupta are with University of California, Los Angeles
E-mail: saptadeep@ucla.edu, puneetg@ucla.edu
}% <-this % stops an unwanted space
\thanks{Manuscript received Oct 3, 2022.}}

\maketitle
\begin{abstract}
Over the past decade, machine learning model complexity has grown at an extraordinary rate, as has the scale of the systems training such large models. 
However there is an alarmingly low hardware utilization (5-20\%) in large scale AI systems. The low system utilization is a cumulative effect of minor losses across different layers of the stack, exacerbated by the disconnect between engineers designing different layers spanning across different industries. 
 %What is needed is a framework to connect researchers/developers across different layers of the stack to communicate their needs, and understand the impact of their design decisions trickling up and down the stack.
We propose CrossFlow, a novel framework that enables cross-layer analysis all the way from the technology layer to the algorithmic layer. 
We also propose DeepFlow (built on top of CrossFlow using machine learning techniques) to automate the design space exploration and co-optimization across different layers of the stack. We have validated CrossFlow accuracy with distributed training on real commercial hardware and showcase several DeepFlow case studies demonstrating pitfalls of not optimizing across the technology-hardware-software stack for what is likely, the most important workload driving large development investments in all aspects of computing stack.   
\end{abstract}
%]

\section{Introduction}

Over the last decade, the demand on compute and memory resources for AI workloads has grown by multiple orders of magnitude~\cite{openai}.
%Over the last decade, the amount of compute and memory for AI has grown by multiple orders of magnitude~\cite{openai}. 
%The exponential growth in computation demand coupled with the slowdown of Moore's law
%have shifted the AI hardware landscape towards specialization and distributed scale-out computation.
As AI models grow in size along with the volume of training data, distributed training on cutting-edge scale-out systems composed of a large number of accelerators and processors has become the norm.
%However, there is a growing concern over low-utilization of such large-scale systems.
However, it has often been noticed that large scale AI training suffers from poor resource utilization. E.g., recent analysis reveals 5-20\% utilization across 1000s of GPUs~\cite{KunleScaledML}. Such poor utilization of resources is becoming a source of major concern.
Inefficiencies across different layers of the compute stack~\cite{jia2018beyond,aws} (from hardware micro-architecture to software parallelization strategies) and the design imbalance across different layers are among a few factors resulting in such low system utilization.
Different layers of the stack, technology nodes, hardware architecture, network topology, model architecture, parallelism strategy are designed across different organizations and retrofitted into the large-scale systems. The distributed nature of the design makes cross-layer optimization challenging if not impossible. 
For example, high-level design decisions like batch size, model architecture, and parallelism strategy exploited at algorithmic level
stress underlying hardware components (network, memory bandwidth or compute units) in different ways which call for different architectural designs, network topologies, memory technologies and technology nodes to ensure high system utilization.

Despite this, the distributed AI training hardware landscape often focuses on just a small set of parallelism strategies for a fixed hardware design~\cite{jia2018beyond}. Exploring the trade-offs between parallelization strategy (e.g. data parallelism and model parallelism) and performance (run-time) is often done in an ad-hoc manner. 
There is no methodical framework or research that explores the trade-offs between low-level hardware technology details and high-level algorithmic design (such as model architecture, parallelism strategy and batch size) on over performance and utilization of compute and memory resources. % under a fixed area and power budget.
%
%In this work, our goal is to develop a framework that enables cross-stack optimization/exploration analysis.
As a result, we set out to develop a framework that could enable across-the-stack analysis and allow us to look at the optimal points in the vast technology, system and algorithm design space. Towards that goal, we develop \textbf{CrossFlow}, a performance modeling framework that enables ``what-if'' analysis across different layers of the stack, and \textbf{DeepFlow} that builds on top of \textit{CrossFlow} and uses machine-learning based techniques to automate the design space search. %\footnote{We will open-source and publicly release the tool.}
CrossFlow is an end-to-end performance modeling tool based on an analytical model which takes the entire system-architecture into account and is more sophisticated than a simple Roofline analysis and less time-consuming than simulation. 
The framework provides a templatized interface for defining technology (minimum operating voltage, bitcell area, etc), chip (compute cores, memory hierarchy, etc.) and system-level architecture (node-level organization, intra-node network, and inter-node network), machine-learning model's compute graph, and parallelization strategies and predicts run-time per iteration step. Key contributions of this work include:
\begin{itemize}
    \item We conduct a variety of case studies looking at impact of a variety of high-cost technology innovations on eventual performance of distributed  DL training. We show that future logic technology nodes alone would provide minimal performance gains, and advancement in HBM and inter-node network technologies is needed to provide the next leap in performance. Also, optimal parallelism strategy selection could provide more performance gains than using naive parallelism strategies on next generation hardware (Section~\ref{sec:case-study}).
    \item We develop the first open-source, full-stack pathfinding framework, \href{https://github.com/nanocad-lab/DeepFlow}{DeepFlow}\footnote{\url{https://github.com/nanocad-lab/DeepFlow}}, for large distributed deep  learning (DL) training: the driving workload for most future technology, hardware and software development (Section~\ref{sec:overview}-~\ref{sec:dse}).
    \item We validate CrossFlow performance prediction against measurements on real commercial hardware (NVIDIA P4, V100 and DGX-1) running kernels and DL application in both single and distributed settings, observing near perfect correlation and 10\% - 16\% error.
    %and show that many of them may not deliver on the promise. 
    Next we show that large multi-chip  integration and waferscale technologies would not be worthy investments for large scale language models (Section~\ref{sec:validation}).
\end{itemize}

CrossFlow and DeepFlow can be used to bridge researchers across different layers of the stack (often spanning across different industries) to communicate their needs. 
\begin{figure}[tbp]
\centering
\begin{subfigure}
  \centering
  \includegraphics[width=0.45\linewidth]{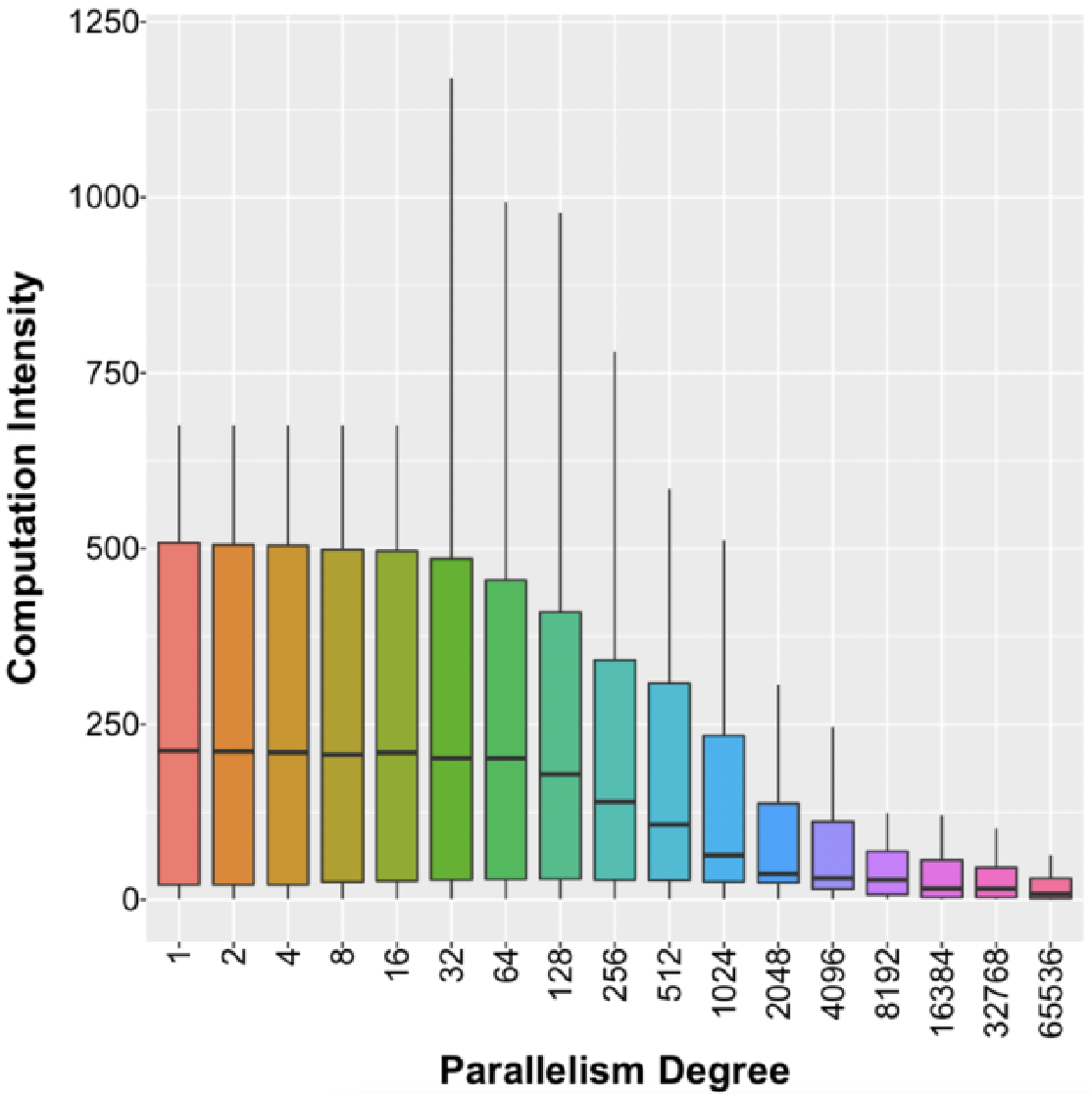}
  %\caption{Parallelism Degree}
  \label{fig:compint1}
\end{subfigure}%
\begin{subfigure}
  \centering
  \includegraphics[width=0.45\linewidth]{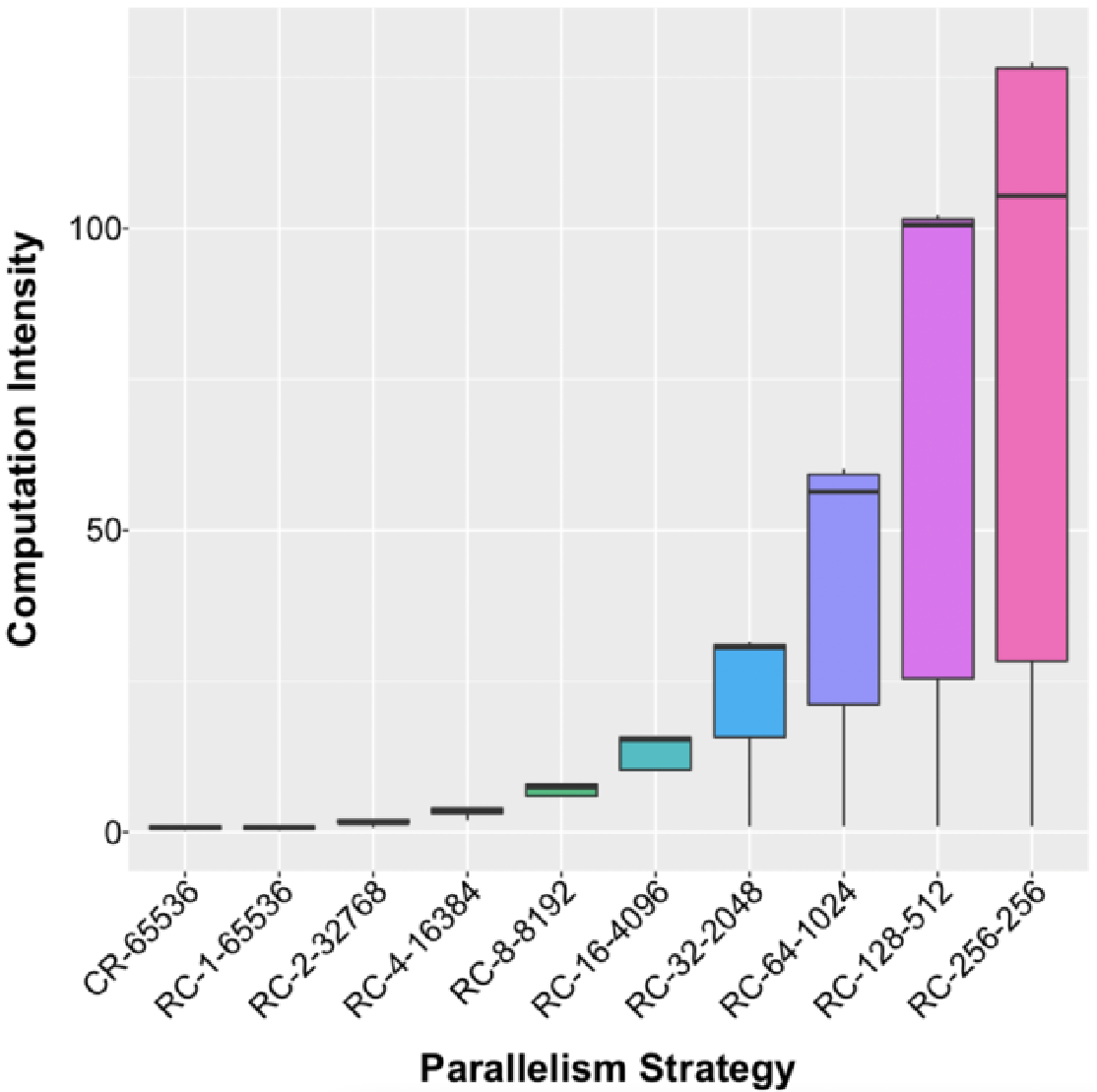}
  %\caption{Parallelism Strategy}
  \label{fig:compint64k}
\end{subfigure}
%\begin{subfigure}
%  \centering
%  \includegraphics[width=0.32\linewidth]{figs/figs/commint.png}
%  %\caption{Parallelism Strategy}
%  \label{fig:commint}
%\end{subfigure}%
\caption{Impact of Parallelism on Computation Intensity.}
\label{fig:compint}
\end{figure}
\vspace{-0.1cm}
\section{Motivation}
High-level algorithmic design decisions such as batch size, parallelism strategy and degrees of parallelism stress the underlying hardware components in different ways. %To maximize the underlying hardware utilization, different hardware designs are required. 
One important metric that guides a balanced compute-memory design is computation intensity. Computation intensity is a workload property defined as the ratio of the number of computation flops to number of accesses to main memory.
%This ratio dictates the optimal ratio of computation throughput to memory bandwidth in the underlying hardware accelerator.

Figure~\ref{fig:compint} (left) shows the computation intensity distribution across different number of GPUs. We performed this analysis for a GEMM problem of size $(64K, 64K, 64K)$ distributed across many GPUs. Depending on the parallelism strategy and number of available GPUs, each GPU gets a non-regular matrix shards for compute.
Each boxplot shows the spread of computation intensity for different number of GPUs.
For each level of parallelism, we see a large spread of compute intensities, particularly for lower parallelism degrees. This is the result of different parallelization strategies as well as different tiling strategies.
It is clear from this figure that computation intensity is much smaller at higher degrees of parallelism, implying the need for a different design point. 

%Besides the parallelism degree, the choice of parallelization strategy has a direct impact on the utilization of the underlying hardware.
There are a myriad of ways to parallelize a model across a large multi-node system.
Figure~\ref{fig:compint}(right) shows the distribution of computation intensity across different parallelization strategies for a fixed level of parallelism (64K GPUs). 
On the X-axis, we show various parallelization strategies across 64K GPUs. 
RC or CR refers to Row-Column or Column-Row distributed GEMM (a.k.a kernel parallelism, more details in Section~\ref{subsec:par_strategy}).
%It is clear from the figure that computation intensity is different across different parallelization strategies,  implying 
As shown, optimal design point is different for different parallelization strategies.
%%High-level parallelism not only controls the design of the hardware nodes but also the network connecting them together. 
%One metric that guides the optimal design point at network level is the ratio of the number of bytes transferred to far memory over network to near memory. 
%We refer to this metric as communication intensity. Figure~\ref{fig:comint}(bottom) shows the distribution of communication intensity across different degrees of parallelism.
%
%Parallelization not only influences the design of the accelerator node but also the network infrastructure that connects them together. 
%This includes network topology and network bandwidth.
%For a given network topology, the metric that guides the bandwidth decision is the amount of data bytes transferred from one node to the next. 
%The data bytes transferred over the network from one node to the next vary by levels of parallelism and the type of parallelism strategy (RC vs CR).
%
%Therefore, the optimal hardware architecture strongly depends on high-level algorithmic and software decisions. Hence, an exploration framework is needed that would allow one to obtain and analyze the various possible optimal system and algorithm design combinations. Furthermore, 
%
%These results indicate dependency between high-level algorithmic design decisions and low-level hardware design, hence the need for a cross-stack co-design. 

Large training workloads are rapidly becoming the applications driving large investments in semiconductor technology development all the way down to fabrication equipment, making such a cross-layer pathfinding framework immensely valuable to ML engineers, system architects and technology developers alike. 

\begin{figure*}[htbp]
\centering
\captionsetup{justification=centering}
\centerline{\includegraphics[width=0.98\linewidth]{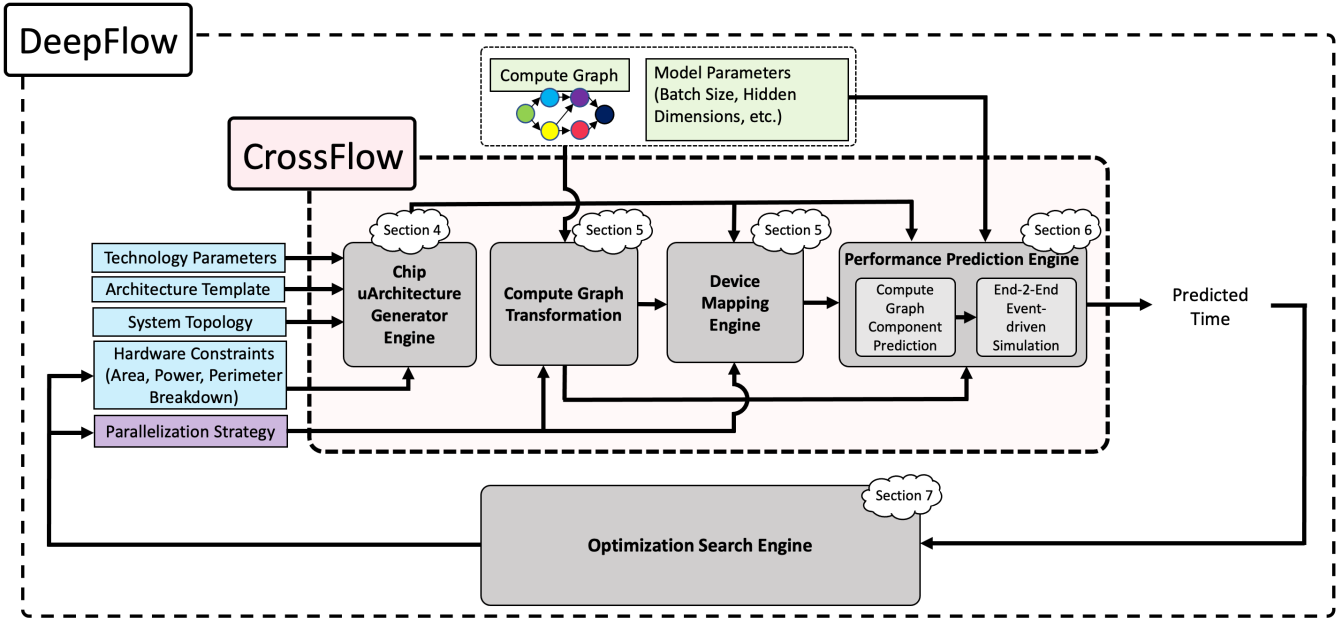}}
\caption{DeepFlow Overview.}
\label{fig:overview}
%\vspace{-0.5cm}
\end{figure*}
%gemm_val.tex
%\vspace{-0.4cm}
\section{DeepFlow Overview}\label{sec:overview}

Figure~\ref{fig:overview} shows an overview of the \name framework. \name takes the following set of \textbf{inputs}: 
(1) \underline{System} design hierarchy (e.g., the number of accelerator nodes per device, the number of devices in the system, the network topology connecting nodes within a device and across the devices), 
(2) \underline{Architecture template} of each accelerator node which provides a high-level definition of its components and how those components fit together. The purpose of the template is to provide a blueprint for the accelerator without committing to any specific hardware parameters.
%A component definition (e.g., minimal compute units (MCU\footnote{Examples of what we regard as MCU includes SMU in older GPUs, Tensor Cores in newer GPUs or systolic array in TPUs}), memory hierarchy, network), specification of each component (e.g., flop rate for each MCU, MCU dimensions, number of MCUs sharing a set of register files, dataflow execution model, and characteristics and scope of different levels of memory hierarchy), 
(3) \underline{Technology} parameters for each hardware component (e.g. energy per flop), 
(4) \underline{Design budgets} for each hardware component (area, power, perimeter),  
(5) \underline{Machine learning model} specification in the form of a high-level compute graph, parameters of each compute node (kernel type, tensor dimensions), and
(6) \underline{Parallelism strategy} (data, model, kernel, and/or pipeline parallelism dimensions) which distributes the compute graph across the entire system. 
(7) \underline{Device mapping} strategy which defines mapping of parallel shards onto hardware nodes.
Given these inputs, \name predicts the end-to-end performance of one iteration (i.e., single batch) of the model and finds an optimal hardware-software-technology design point as \textbf{output}. 

DeepFlow is composed of two major components.
\underline{CrossFlow} which operates in a stand-alone mode and can predict performance for any input configuration; and a search and optimization engine (\underline{SOE}) which enables design space search. 
%To do so, \name breaks the problem into multiple phases.
%Each phase or building block of \name is described in details next.
\vspace{-0.1cm}
\subsection{CrossFlow Building Blocks}

\paragraph*{\em Micro-Architecture Generator Engine (AGE)}

AGE takes the following set of \textbf{inputs}:
(1) Design constraints (i.e the power, area and perimeter budget and breakdown across micro-architectural components such as cache, network, compute cores). 
This breakdown can be provided manually by users or automatically by the Search and Optimization Engine (SOE, Section~\ref{subsec:soe}).
%We also provide technology specifications such as 
%and their physical characteristics such as area/power per core under nominal operating conditions, SRAM/register characteristics. 
(2) Technology parameters such as energy per flop, energy per data bit transfer for each level of memory and network hierarchy, threshold and maximum gate voltage, integration substrate parameters such as bump/interconnect pitch. We provide a wide range of standard and future technology libraries as baseline. (3) Architecture template which is a blueprint of the underlying accelerator chip without committing to any specific hardware parameters. Given these input, AGE performs a frequency-voltage-area scaling optimization to generate the following \textbf{output} parameters such that design budgets for all component are met: 
(1) Compute throughput.
(2) Capacity for different levels of memory hierarchy.
(3) Bandwidth to each level of memory hierarchy.
(4) Inter-node as well as intra-node network bandwidth. 
These parameters are then utilized by the performance prediction engine (PPE) to estimate the execution time of each kernel.
%As mentioned previously, 
%The output of this stage is the input to performance engine to estimate the execution time of each kernel. Next, we describe the search and optimization engine (SOE) which feeds input values to AGE, if we want to use the model for architecture search.
%\vspace{-0.2cm}
\paragraph*{\em Compute Graph Transformation and Device Placement Engine (DPE)}
The parallelization strategy and device mapping are critical in deciding the overall execution time. Here, we first transform the model graph to a `super-graph' to reflect the parallelization strategy provided by the users manually, or SOE engine (Section~\ref{subsec:soe}) automatically. For example, to apply data parallelism, the model graph is replicated and appropriate edges are added to model the gradient exchange. After generating the transformed graph, DPE assigns the vertices of the transformed graph to the system nodes following a heuristic approach to minimize the communication overhead. %
%The details are presented in section~\ref{}.

%\vspace{-0.2cm}
\paragraph*{\em Performance Prediction Engine (PPE)}
%With the device mapping for all the vertices of the compute (super-)graph known, the next step is to calculate the overall execution time for a forward pass and/or a backward pass. 
We use hierarchical roofline modeling to predict the performance of each compute node. To calculate the overall end-to-end execution time, while respecting scheduling constraints (e.g. one kernel at a time per GPU, or prioritizing one kernel launch over another) we use event-driven simulation.%
%We explain the details of the PPE in section~\ref{}.
\subsection{Search and Optimization Engine (SOE)}\label{subsec:soe}
Co-optimizing micro-architectural parameters and the parallelization strategy that minimizes the overall end-to-end execution time requires navigating a large space of design parameters. 
Search and optimization engine (SOE) enables the automatic design space search and finds an 
%that meets the total power and area constraints, and simultaneously explores software parallelization strategies to find the 
optimal design point which meets the design constraints and minimizes the overall execution time.
%Because the hardware configuration space is very large, the search algorithm we designed 
SOE takes inspiration from ML-assisted search algorithms, in particular gradient decent search with momentum and builds on top of the CrossFlow modeling engine.
%The software parallelization design space is much smaller compared to the hardware design space and therefore we employ an exhaustive grid search. 

%Gradient search is an iterative process. In each step, SOE takes the predicted time from previous iteration as input to re-adjust the following parameter settings: (1) power, area and perimeter breakdown across different architectural components. (2) a parallelization strategy. These parameters will be fed back to CrossFlow to estimate the overall execution time. This process continues until convergence or user-specified number of steps. 
%The details of SOE's search algorithm are elaborated in Section~\ref{}. 
\vspace{-0.2cm}
\subsection{Parallelism Strategy Space}
\label{subsec:par_strategy}
There are a myriad of ways to parallelize a model across a large multi-node system. Exploring the parallelism space and finding the optimal strategy is critical to overall performance and system utilization. DeepFlow explores kernel, data and layer parallelism. It uniquely identifies each parallelism strategy by following notations: $\texttt{RC-\{KP1\}-\{KP2\}-d\{DP\}-p\{LP\}}$ or $\texttt{CR-\{KP1\}-d\{DP\}-p\{LP\}}$ depending on the choice of kernel parallelism.
RC (Row-Column) and CR (Column-Row) refer to different forms of kernel parallelism, i.e. distributed GEMM through inner-product or outer-product implementation.
%\begin{equation*}
%    \texttt{RC: R{KP1\}\_C\{KP2\}\_d\{DP\}\_p\{LP\}}
%\end{equation*}
%Where \texttt{RC} or \texttt{CR} refers to the type of kernel parallelism strategy, i.e. Row-Column or Column-Row,
%\texttt{N} refers to the number of parallel nodes,
\texttt{KP1} and \texttt{KP2} are the parameters of distributed GEMM. 
For Row-Column (\texttt{RC}) or inner-product, \texttt{KP1} and \texttt{KP2} would refer to the number of ways we shard the first matrix across rows and the second matrix across columns.
For Column-Row (\texttt{CR}) or outer-product, we would only need one parameter to specify the parallelization strategy; \texttt{KP1} will refer to the number of ways we cut the first matrix across columns and the second matrix across rows.
\texttt{DP} represents the number of model replicas and data shards assigned to each to exploit data parallelism.
\texttt{LP} is the number of ways we cut layers into stages to exploit pipeline parallelism.

\section{Micro-Architecture Generator Engine}
The micro-architecture generator engine, AGE,  takes three sets of inputs: (1) A technology components library, where the characteristics of each component such as cores, different types of memories, network interfaces, etc. are defined, (2) Architecture template, where the overall high-level chip and system organization (such as compute and memory hierarchies) is provided, (3) Hardware resource allocation, where area, power, and chip perimeter budgets are provided for the different components of the system. Using this information, the AGE generates the final micro-architecture parameters (such as overall compute throughput, memory bandwidths at different memory levels, network bandwidth) as shown in Figure~\ref{fig:overview}. 

%In this section, we will first provide details of each of the three inputs and then describe how the uArchitecture generator engine uses these inputs to generate the final uArchitectural parameters of the system.

\subsection{Technology Components Library}

A system is generally composed of many primitive components or building blocks such as the compute units, SRAM banks, DRAM, interconnect network components (on-chip and off-chip), etc. A library of these components and their associated technology parameters are provided as input to the tool through a $tech\_config$ YAML file. We classify these components into three primary categories: compute, memory and network. 
%Next we describe the different hardware attributes we assign to the components in each category.
%\vspace{-0.1in}
\subsubsection{Compute}  
Attributes for the minimal compute components such as matrix-multiplier units, vector-matrix multiply units, or a dataflow architecture unit like systolic array are specified under this category. When a compute component is added to the library, the compute attributes listed in Table~\ref{tab:component-parameter} will have to be defined for that component. The tool user can add any type of compute component in the library ranging from a simple scalar unit to a complex unit comprising of a bundle of systolic arrays and capture the micro-architectural characteristics in the final architecture template file.
%\vspace{-0.1in}
\subsubsection{Memory}
The memory components in a system can be built out of different technologies (e.g., SRAM, DRAM, MRAM, RRAM, 3D-XPoint). Also, these memory components can be used in two ways: on-chip memory and off-chip memory. A library of fine-grained memory components can be created and stored under this category which is utilized to construct different levels of the memory hierarchy. The characteristics of the on-chip components are described at the granularity of a bank because the smallest on-chip memory unit available to a system designer is usually a memory bank. The parameters of a memory bank such as capacity, bit area, periphery overhead etc. are taken as inputs. 
On the other hand, we model the off-chip memory components such as DRAM, or 3D-XPoint at device level granularity, e.g., an HBM stack. This is because the off-chip components are usually obtained at a device level granularity. For off-chip memories, other parameters such as memory controller area, I/O bus width per device, etc. need to be defined. This information is then used to precisely model the capacity and throughput of different levels of the memory hierarchy under the given area and power constraints.
%\vspace{-0.1in}
\subsubsection{Network}
The inter-chip network component is  either intra-node or inter-node communication link. In the case of a multi-chip module (MCM) where multiple compute dies and memory devices are integrated on a 2.5D integration substrate within the same package, the inter-die communication is done using high density and energy-efficient links on the 2.5D substrate. These links are considered as intra-node links. On the other hand, the off-package communication links between nodes are considered as inter-node links. The attributes that need to be defined for inter and intra-die communication network components are provided in Table~\ref{tab:component-parameter}. In case of a waferscale system, the entire wafer could be considered as a single node.

\begin{table}
\centering
\resizebox{\linewidth}{!}{%
\begin{tabular}{|c|l|l|}
\hline
\multirow{4}{*}{\textbf{Compute}} & Technology Node & Nominal Area \\ \cline{2-3} 
 & Nominal Voltage & Threshold Voltage \\ \cline{2-3} 
 & Nominal Frequency & Minimum Voltage \\ \cline{2-3} 
 & Nominal OP rate & Maximum Voltage \\ \hline
\multirow{4}{*}{\textbf{On-chip Memory}} & Technology & Latency \\ \cline{2-3} 
 & Dynamic energy per bit & Static energy per bit \\ \cline{2-3} 
 & Area per bit and total area overhead & Bank Capacity \\ \cline{2-3} 
 & Controller area overhead per bank & Controller power overhead per bank \\ \hline
\multirow{6}{*}{\textbf{Off-chip Memory}} & Technology & Number of links per device \\ \cline{2-3} 
 & Dynamic energy per bit & Nominal Voltage \\ \cline{2-3} 
 & Static power per bit & Nominal Frequency \\ \cline{2-3} 
 & Device Capacity & Minimum Voltage \\ \cline{2-3} 
 & Device Area & Maximum Voltage \\ \cline{2-3} 
 & Memory Controller and I/O Area & Access Latency \\ \hline
\multirow{4}{*}{\textbf{\begin{tabular}[c]{@{}c@{}}Network (intra-node \\      and inter-node)\end{tabular}}} & Nominal Voltage & Number of links per mm \\ \cline{2-3} 
 & Nominal Frequency & Threshold Voltage \\ \cline{2-3} 
 & Nominal Energy per Link & Minimum Voltage \\ \cline{2-3} 
 & Nominal Area per Link & Link Latency \\ \hline
\end{tabular}%
}
\caption{Different technology components.} \label{tab:component-parameter}
%\vspace{-0.2cm}
\end{table}

\begin{figure}[t]
%\vspace{-0.4cm}
\centerline{\includegraphics[width=0.9\linewidth]{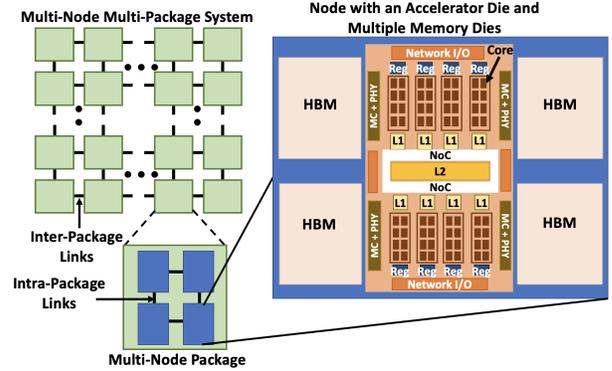}}
\caption{{\textbf{Architecture Template}: Overview of a hardware system whose characteristics can be configured in DeepFlow.}}
\label{fig:hardware-system-overview}
\vspace{-0.1cm}
\end{figure}
\subsection{Architecture Template}
Once all  system components are instantiated from the technology library, the next step is to hierarchically organize one or multiple components from each category to construct the overall system.  Distributed machine learning training is done on scale-out multi-node system, as shown in Figure~\ref{fig:hardware-system-overview}. Such a system consists of multiple individually packaged nodes which communicate through off-package interconnects (such as NVLink, Infiniband etc.) that form the inter-node network. Inside each package, there can be multiple different accelerator nodes connected using an intra-node network. Each accelerator within the package typically consists of one accelerator die that is connected to its own off-chip main memory components (such as HBM, as shown in the figure). Each accelerator die itself can be composed of smaller compute units.

\name  provides a rich template that can be used to specify the overall architectural organization of such an accelerator system. 
%The template is used to specify the high-level micro-architectural organization of the compute units, organization of the memory hierarchy, inter-chip network topology (both intra- and inter- node) and the system hierarchy. 
Next we describe in detail how the template is organized and how different system configurations can be achieved using this template.

\subsubsection{\textbf{Compute unit}}\label{subsec:mcu} As shown in the accelerator die architecture in Figure~\ref{fig:hardware-system-overview}, compute units are often organized in hierarchies. E.g., in an NVIDIA GPU, multiple tensor cores are bundled in a streaming multi-processor (SM) and the SM as a whole interacts with the cache hierarchy. In \name one can express such hierarchy by defining \textit{minimal compute units (MCUs)} and
\textit{MCU bundle}. MCU is the smallest compute unit that we expose to the tool user.  It defines the dataflow model and layout (e.g. MCU can be a systolic array that its height and width are configurable as input) and interacts with the first level of memory hierarchy. Meanwhile, MCU bundle defines the number of MCUs that are bundled together and are exposed to the second level of memory hierarchy. 

In dataflow architectures such as Eyeriss, TPU etc., data can flow directly between different cores. Hence, the tool allows one to define the type of dataflow within an MCU. Currently the performance model supports three types of dataflow: \textit{weight stationary, activation stationary and output stationary}. The tool can also find the best dataflow strategy among the three for any given kernel.% if the dataflow mode is selected as `best'.

Software runtime, scheduling overheads and the architecture of the cores often restrict the maximum compute utilization. For example, the tensor-cores in NVIDIA V100 incurs fill-drain related under-utilization during tensor loading from the registers and therefore achieves a maximum utilization of 85\%. To account for such overheads, a maximum utilization value can be defined which derates the core throughput by that factor. 

\subsubsection{\textbf{Memory Hierarchy and Scope}} The memory hierarchy is defined by initializing multiple memory levels from the highest to the lowest level (i.e., registers to the main memory) as shown in Figure~\ref{fig:hardware-system-overview}. Each level of memory has two attributes: (1) \underline{Memory technology} component from the technology component library which defines the physical attributes of the memory as outlined in Table~\ref{tab:component-parameter}, and (2) \underline{Scope} defines 
the set of components from the next level of memory hierarchy that are accessible from this level of memory hierarchy.
%which micro-architectural components can access that level of memory. 
For example, the `global' scope indicates that the memory level is accessible to all the components.

\subsubsection{\textbf{Network Topology}} In \name, we support two levels of network hierarchy: intra-package and inter-package. For each level, a different topology (e.g. mesh, torus, crossbar) can be defined.

\subsection{Hardware Resource Allocation}
 Hardware design under a limited area and power budget is a fine art of finding the right balance (breakdown of resources) across different micro-architectural components. 
 The area and power allocation for each micro-architectural component, as well as the perimeter allocation for certain components derive the design and specification of that component. 
 
We define resource (area, power, perimeter) distribution across different components of the compute chip, as input parameters.
The input definition also includes the total area and power budgets for the entire compute node. The total perimeter is inferred from area. 
The area budget is usually dictated by packaging constraints. For example, if the compute and memory dies are assembled on a 2.5D silicon interposer-based interconnect substrate, the total area of the node will be limited by the maximum size of the interconnect substrate that can be fabricated. 
Compare this to a waferscale system which houses an entire node on a wafer where the total area budget can be as large as 70,000 $mm^2$. 
A node's power budget is determined by the cooling infrastructure that extract heat from the node and the power delivery constraints.

We define budget distribution across different components of the compute graph as a percentage breakdown.
As shown in the YAML snippet
in Figure~\ref{fig:area-breakdown}, 
fractions of the total area is distributed across cores, levels of memory hierarchy and network
components. Similarly, the fraction of the compute chip's power and perimeter gets devoted to different hardware components.
\begin{figure}
\begin{minted}[
    breaklines,
    frame=single,
    fontsize=\tiny
  ]{yaml}
area_breakdown:
    node_area_budget: 1230 #mm2
    proc_chip_area_budget: 815 #mm2
    core: 0.35
    L2: 0.14
    L1: 0.1 
    L0: 0.2
    DRAM: 0.05
    network:
      intra_package: 0.06
      inter_package: 0.1
\end{minted}
\caption{Resource breakdown example: This example is showing the area budget allocation and breakdown across all micro-architectural components.}
\label{fig:area-breakdown}
%\vspace{-0.3cm}
\end{figure}

Given the overall resource allocation and distribution, the AGE performs a series of optimizations (voltage-frequency scaling) to find an optimal parameter settings for each micro-architectural component. An optimal parameter setting is one that utilizes the most of the allocated budget. Note that an unbalanced resource allocation may leave some of the budget under-utilized. While we allow users to provide a manual breakdown of resources as input, we highly recommend to use SOE (Search and Optimization Engine) to automatically find the best setting which maximizes the overall resource utilization. 
%then calculates the micro-architectural parameters of various chip components which would be used by the PPR for performance prediction. 
%Examples of such output parameters include compute throughput, capacity and bandwidth to the different memory levels and the intra- and inter-node network bandwidths. 

%Often at early stages of an accelerator design, designers have a rough idea of resource budget for each u-architectural component but not necessarily the final parameters.
%This stage mimics the real-world process of hardware design.
%Next, we elaborate the micro-architectural paramter generation for each of these components in detail.

%optimally allocate resources corresponding to each component provided in the architecture template. 
%Besides, the perimeter of the compute die determines the total number of wires that can escape out of the compute chip. Hence, the tool also takes as input the breakdown of the perimeter of the compute die to allocate portions of die perimeter to different off-chip links such as off-chip memory links, intra-node and inter-node links.

\subsection{Micro-architectural Parameter Generation}
%\textcolor{blue}{Can we potentially move this to appendix if crunch for space?}
Next, the tool generates the micro-architectural parameters for each component of the architecture. Given the architecture template, alongside the resource breakdown across the different components, and the technology parameters, we find the maximum throughput for each component. E.g., We find the maximum number of cores that can fit in the given area allocation and find the voltage-frequency points to maximize compute throughput under the power budget. Similarly, for on-chip caches, we find the memory capacity and memory bandwidth at each level that can fit in the area budget while taking the network and controller overhead into account. For off-chip memories and network interfaces, we use the energy per bit information along with the physical I/O transceiver  area, bump pitch as well interconnect wiring pitch to determine the maximum bandwidth that can be realized on the chip (using a model similar to \cite{C111}).

These architectural parameters, throughput, bandwidth, capacity etc., are then provided as input to the performance prediction engine. Next, we discuss in detail how we model and calculate these parameters.

%More details about how we model and calculate these parameters have been explained in Appendix~\ref{detailed_uarch}

\subsubsection{Core} For deep learning models, the kernels are usually highly parallel in nature and therefore, our goal is to maximize total compute throughput under the area and power budgets allocated for compute. Given the area budget, we first compute the maximum number of MCUs (minimal compute units, introduced in Section~\ref{subsec:mcu}) that can fit within the area allocated. 
%Then, we compute the voltage and frequency of the compute units such that the total peak power is within the power budget allocated. 
The nominal frequency and voltage for each MCU is an input to the model, therefore the nominal power for each MCU and the entire core can be derived very easily. 
If the nominal power exceeds the power budget, we scale down the frequency and voltage. If we hit the minimum voltage limit set in the component description, we reduce the number of MCUs till we satisfy the total power budget allocated to the compute units.
This explains a case where the core design is power-bound and not area-bound.

Once we determine the total number of cores and the frequency of operation, we compute the compute throughput by appropriately scaling the nominal flop rate, as shown in equation~\ref{eqn:compute_throughput}.
%\begin{comment}
\begin{equation}
    \texttt{Throughput} = N \times \texttt{flop}_{nominal}\times {\frac{f_{op}}{f_{nominal}}}
    \label{eqn:compute_throughput}
\end{equation}

where $N$ is the total number of cores, $\texttt{flop}_{nominal}$ is the nominal flop rate of each core, $f_{nominal}$ is the nominal frequency corresponding to the technology node of the core and $f_{op}$ is the final optimal operating frequency. We use standard Voltage-Frequency-Power scaling methodology to obtain the operating voltage and frequency.
%\end{comment}
%\vspace{-0.1cm}
\subsubsection{Register and Cache Memory} The total area and power budgets allocated to each level of on-chip memory is split between the memory banks and 
the network circuitry that connects the memory banks at each level to micro-architectural components at the next level that are under its scope.
%the on-chip network used to communicate between the microarchitectural components under the scope of that level. 
We assume this interconnect to have a crossbar topology. The total number of components under its scope and the number of banks in that memory level determine the area and power overheads of the network. We iteratively determine the total number of  banks possible at each level of memory hierarchy such that the total area of the banks and the network at every level satisfies the area budget allocation. 
Once we determine the number of memory banks, we calculate total static power of all the banks (Equation~\ref{eqn:static_power_cache}) and we allocate the remaining power budget to dynamic access energy. The available dynamic energy budget determines the maximum achievable throughput as shown in Equation~\ref{eqn:throughput_memory}.
%\begin{comment}

\begin{equation}
    P_{static} = P_{static-per-bit} \times N_{banks} \times \texttt{Capacity}_{bank}
    \label{eqn:static_power_cache}
\end{equation}

\begin{equation}
    \texttt{Throughput} = \frac {P_{on-chip-mem} - P_{static}}{\texttt{Energy}_{dyn-per-bit}}
    \label{eqn:throughput_memory}
\end{equation}
%\end{comment}

%\vspace{-0.2cm}
\subsubsection{Main Memory} 
Main memory has two major components that collectively control the overall capacity and bandwidth but are housed in two different places. Memory controller which is placed on the compute chip, and the memory devices are placed outside the compute die within the same package.
The area allocation to each component determines the maximum number of memory devices that can be supported, which in turn determines the total memory capacity (see Equation~\ref{eqn:num_mem_devices}). 

\begin{equation}
\small
\begin{split}
    \#Devices = min(&\frac{Node\ Area - Processor\ Chip\ Area}{Device Area}, \\ & \frac{Area\ budget\ for\ Memory\ Controllers}{Memory\ Controller\ Area},
    \\ & \frac{Perimeter \times \#Links\ per\ mm}{\#Links\ per\ device})
    \label{eqn:num_mem_devices}
\end{split}
\end{equation}

Meanwhile power and perimeter allocation dictates the number of links (that can fit along the compute die), and the frequency of each link which collectively determine the overall off-chip memory bandwidth. 

\subsubsection{Network} 
The off-chip network links (intra and inter-package) consume both power and area on the compute die. Moreover, the wires need to escape the periphery of the die which gets determined by the interconnect density and the available chip perimeter. The maximum number of links that can be accommodated in the compute die is limited either by the area available to fit in the link I/O cells or the amount of perimeter available for the links to escape the die periphery. Therefore, the tool uses the area per link, the available area budget, wiring density and the die perimeter budget to find the maximum number of links that can fit in the chip. Next, the tool uses the standard voltage-frequency scaling methodology to find the operating point for each link such that the total network-related power is within the power budget allocated. The network bandwidth is then calculated by multiplying the total number of links and the operating frequency of each link. We perform this step for the intra-node network and inter-node network separately.
\begin{figure*}[ht]
\centering
\includegraphics[width=0.9\linewidth]{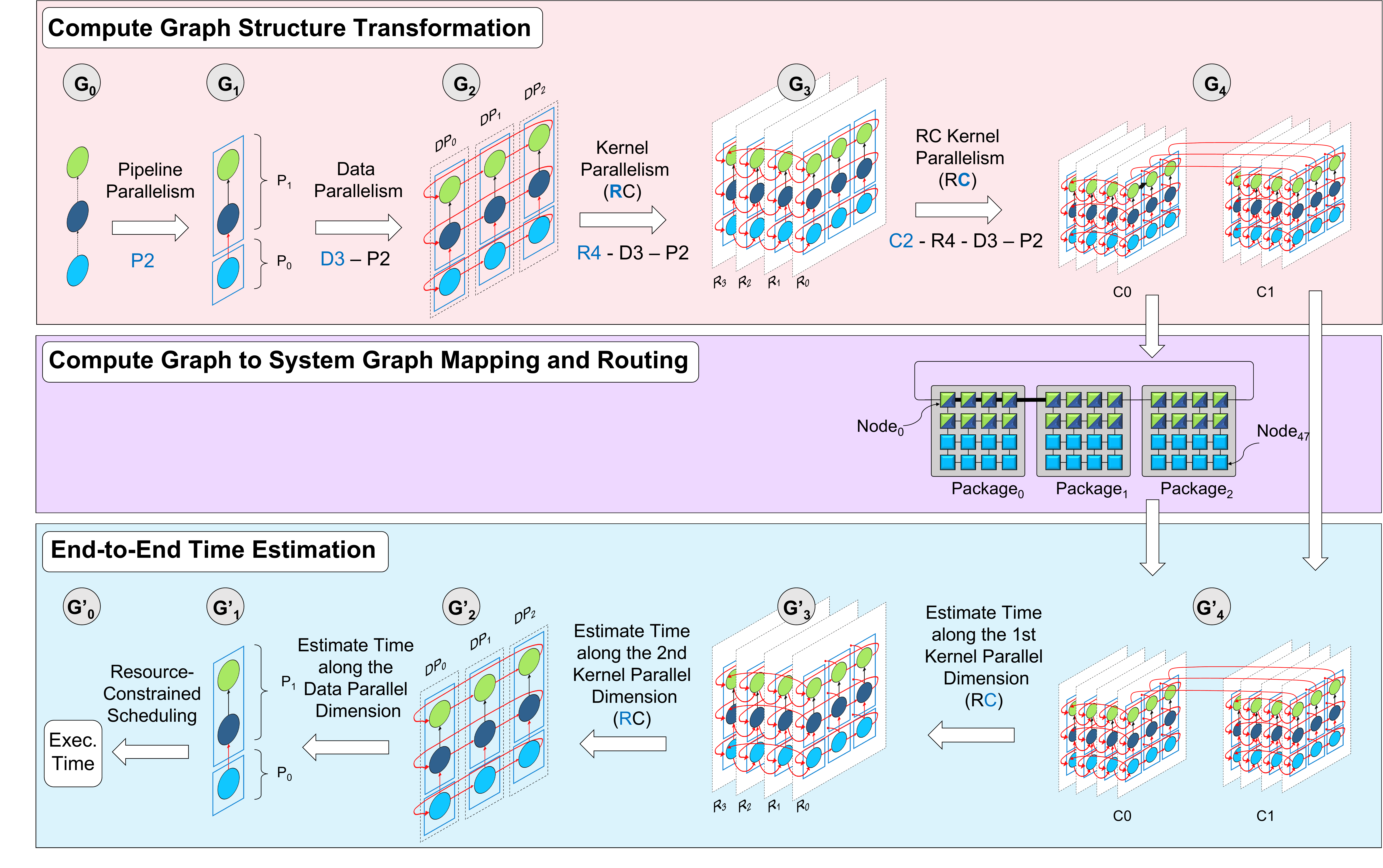}
\caption{\small \textbf{ An Example of a Compute Graph Transformation, Device Mapping and Routing, and End-to-End Time Estimation: (top)} Cross-edges are shown in red. To preserve readability, we only show a subset of cross-edges for kernel parallelism. 
Blue solid borderlines indicates separate hardware nodes. At every parallelization stage, we use black hashed lines to show graph replication along that dimension. A replica is a graph with a similar structure, however the kernel size and/or data size could be different for each replica.
For simplicity, the original graph is a simple 3-layer feed-forward neural network that is divided into two sub-graphs (P2).
Then for each pipeline stage, batch size is distributed across three workers (D3).
Then for each data shard of each pipeline stage, the kernels are distributed in a row-column fashion across a 4$\times$2 torus (RC-K4-K2). \textbf{(middle)} Mapping a 4-D hyper-cube into a 2-D mesh: a greedy layout mapped in the following order: kernel(R), kernel(C), pipeline and data. The bolded black edge in G4 is mapped onto a 4-hop path in the system graph. \textbf{(bottom)} backward pass time estimation. }
%\textcolor{blue}{FIXME: draw a box around the the top row, mark it as graph transformation, draw another box around the bottom row mark it as end-to-end time estimation. Show cross-edges in red.}
\label{fig:transformation}
\vspace{-0.2cm}
\end{figure*}
%gemm_val.tex
\section{Compute Graph Transformation and Device Mapping}
\label{sec:mapping}
%Given the ML model description (in form of a compute graph), the system topology (in form of a system graph) and the parallelization strategy, we use device mapping engine to map the vertices of the compute graph (kernels) onto the vertices of the system graph (nodes) and map the edges of the compute graph (data dependency edges) to the edges of the system graph (physical network links). However, before the mapping happens, we transform the compute graph to reflect the parallelism strategy specified by the user.

%Given the ML model description (in form of a compute graph), and the parallelization strategy, we first transform the compute graph to reflect the parallelism strategy. Next, we use device mapping engine to map the vertices of the compute graph (kernels) onto the vertices of the system graph (accelerator nodes) and map the edges of the compute graph (data dependency edges) to the edges of the system graph (physical network links).

Given the ML model description (in form of a \textit{compute graph}) and the distributed system topology (in form of a \textit{system graph}), 
%an essential step before performance prediction is to 
we find an optimal mapping from vertices and edges in the compute graph to hardware nodes and network links in the system graph. 
%However, there is not a one-to-one mapping from nodes in the compute graph to hardware nodes in the system graph in presence of parallelism: A compute node (kernel) can be replicated and mapped to different hardware nodes (data parallelism), or it can be replaced with smaller sub-kernels (kernel parallelism). This complicates the mapping problem.
However, before mapping, we transform the compute graph into a \textit{super-graph} to reflect the parallelism strategies specified as input.
%to enable a one-on-one mapping between nodes in the compute graph to hardware nodes in the distributed system graph. 

%\vspace{0.1in}
\subsection{Compute Graph Structure Transformation}
%A graph transformation or rewrite defines a set of rules of the form $S \rightarrow R$, with $S$ being the starting graph and $R$ being the result graph. 
%A rewrite rule is applied to the starting graph by searching for an occurrence of the pattern graph (pattern matching, thus solving the subgraph isomorphism problem) followed by replacing the found occurrence by an instance of the replacement graph. 
%A rewrite rule specifies a replacement graph that replaces each node in the starting graph, and how these replacement graphs are connected in the result graph.
%A graph transformation is the stepwise replacement of subgraphs inside a host graph. 
%data parallelism entails replicating the graph $N$ times and connecting the corresponding compute vertices in a fashion consistent with the reduction algorithm: for ring-all-reduce, the corresponding vertices would be connected in a ring fashion.
%A graph transformation or rewrite defines a set of rules that apply to sub-graphs in the \textit{original} graph in order to generate the \textit{result} graph.
%A rewrite rule specifies a \textit{replacement} graph that replaces every instance of a \textit{sub-graph} (pattern matching) in the original graph.
%It also defines how these replacement graphs are connected in the result graph~\cite{}.
Each parallelism strategy is a form of graph transformation where the sub-graph to be replaced is a single node, so essentially all nodes would be replaced with the same replacement graph.
For example, to model data parallelism (with the ring-all-reduce implementation) we would need to 
$\textit{replace}$ each node in the original graph with a ring of length $N$ (for an $N$-data parallel strategy). 
The new edges on the ring will be marked as $\textit{cross-edge}$ to capture the fact that they connect compute nodes hosted on separate devices.
To capture a kernel parallelism strategy (e.g. $\texttt{RC-\{KP1\}-\{KP2\}}$), we would need to $\textit{replace}$ each node in the compute graph with a 2-dimensional torus of $\texttt{KP1} \times \texttt{KP2}$ dimension 
(assuming the reduction algorithm along each dimension is ring-all-reduce). 
Similarly, new edges on the torus would be marked as cross-edge.
To capture a pipeline parallelism, no node transformation is required. The pipeline parallelism slices the original graph into multiple sub-graphs, each hosted on a separate hardware node.
Edges connecting sub-graphs would be marked as cross-edge.
Figure~\ref{fig:transformation} shows the composition of multiple parallelism strategies applied in sequence (pipeline, data and kernel parallelism, respectively). 
$G_0$ is the original compute graph and $G_4$ is the final transformed graph.

%Pipeline parallelism, unlike other forms, does not result in replication or replacement of graph nodes. It simply modifies the edge type between nodes across layers that are mapped to different pipeline stages to capture their cross-connection nature. 

%\vspace{0.1in}
\subsection{Device Mapping and Routing Engine}
%After the transformation stage, the transformed graph would have $N_r\times$ more number of nodes than the original graph, where $ N_r = dp\times kp_1 \times kp_2$, and $dp$, $kp_1$ and $kp_2$ are multiplicative factors of data parallel and kernel parallel strategy. 
%Moreover, to exploit pipeline parallelism, each model replica can be partitioned into $P$ sub-graphs and hosted in separate devices. 
%Data parallelism, kernel parallelism and pipeline parallelism would require each parallel shard to be hosted on a separate physical device. Hence, the total number of hardware nodes, $N_h$, should be $N_r \times P$.
%We use pipeline parallelism to slice the super-graph into smaller partitions such that number of partitions is consistent with total number of replicas and number of physical devices. 
Data parallelism, kernel parallelism and pipeline parallelism would require that each parallel shard to be hosted on a separate physical device.
Hence, device mapping happens at the granularity of a parallel shard. 
%Each parallel shard is basically a sub-graph in the transformed graph.
We want parallel shards that are close in the parallel space to be mapped onto nodes that are close in the physical space to minimize communication. 
%However, parallelism space is usually a 5 or 4 or 3-dimensional hypercube, while the underlying system graph is usually a 3 or 2-dimensional mesh or torus.
However, the transformed graph usually has higher dimension than the system graph. 
Figure~\ref{fig:transformation} shows such example, where the final transformed graph ($G_4$) is 4-D hypercube and the system graph is a 2-D torus.
Therefore, it will not be possible to map all adjacent nodes in the compute graph to adjacent nodes in the system graph.
We adopt a greedy approach to conduct such mappings: We start with a parallel dimension, map all parallel shards along that dimension to adjacent nodes in the hardware.
If the number of shards along the parallel dimension is larger than the hardware dimension we are mapping onto, we wrap-around to the next immediate dimension.
We continue this process along other dimensions in a specific order, until all nodes are mapped.
The order at which we walk along the parallelism dimensions results in different mappings.
For 4 different parallelism strategies, we explore $(4!)$ = 24 possible orderings to pick the best mapping.
Once node mapping is determined, we take a last step to map edges to physical links.
An edge that connects to adjacent node in the compute graph may map to a multi-hop path as shown in Figure~\ref{fig:transformation}. 
As a result, one physical link would be shared across multiple edges.
The number of logical edges sharing a physical link is an important factor for effective bandwidth estimation.
We use $X-Y$ routing to map edges in the compute graph to paths in the system graph.
Overall, the whole transformation step followed by device mapping is necessary to find an accurate estimation of \textit{edge} timing.

%Depending on the parallelism strategy, the transformed compute graph can take the form of a 5D/4D/3D hypercube. However, the underlying system topology may be a 2D or 3D hypercube. This means that multiple edges in the compute graph would need to be mapped in to one edge/link of the system graph. The number of logical edges that are mapped into a physical edge is an important metric as it captures the effect of link bandwidth sharing which ultimately affects the communication overhead. Based on link sharing, we allocate the derated/partial link bandwidth to the edges of the compute graph. In order to find a mapping that minimizes the maximum number of edges that is mapped in to each network link, we use a greedy heuristic. Our heuristic works as follows: starting with one axis in the parallelization space, we map the vertices along that axis as close as possible, before moving to the vertices in the second axis. We repeat this for different permutations of axes and choose the one which minimizes the communication overhead.

%\todosaptadeep{These different parallelism strategies have major implications on end-to-end performance and stress different resources of the system. As an example, very large models often wouldn't fit in the main memory capacity of a single accelerator device, and therefore model and/or pipeline parallelism needs to be employed. Model parallelism on the other hand, can stress inter-node network characteristics.}
%\input{figs/validation}
\section{Performance Prediction Engine}\label{sec:ppe}
%\name performance prediction engine predicts time for each node and each edge within the graph individually. 
%Having time for each node and each edge within the compute graph, we will use an event-driven simulator 
%that walks over the graph and triggers new event dictated by data dependency and resource mapping schedule.
Once mapping is decided for each node and each edge in the transformed graph, performance prediction engine estimates timing for each node and each edge. 
%The end-to-end performance estimation is conducted in a hierarchical manner. 
%We first predict the timing for each node in the transformed graph. 
We then use a resource-constrained scheduling algorithm to find the end-to-end timing. %We explain these steps in more details.

\vspace{-0.2cm}
\subsection{Hierarchical Roofline} 
We use hierarchical roofline analyses~\cite{roofline} to predict the timing of each node in the transformed compute graph. 
%Roofline analysis is an analytical approach that predicts if an application is compute-bound or memory-bound. 
%Once this is known, our tool uses a simple throughput analysis to estimate timing.
%: i.e. $\texttt{\#flops/compute throughput}$ if compute-bound, or $\texttt{\#memory accesses/memory bandwidth}$ if memory-bound.
For each node, we estimate the operational intensity ($\texttt{OI$_L$ = \#flops/\#memory accesses$_L$}$) to each level in the memory hierarchy.
%
%For systems with multiple levels of memory hierarchy, we adopt a hierarchical roofline analysis.
%Hierarchical roofline predicts if an application is compute-bound, L1-bound, L2-bound, memory-bound, etc.
We search over the space of possible tiling strategies at each level of memory hierarchy and estimate the number of memory accesses to each level. We explain this in more detail next.
%which in turn rely on accurate estimation of number accesses to each level. 
%Operational intensity is different for different levels of memory hierarchy, determined by the number of accesses to each level. Hence, our accuracy very much depends on accurate estimation of accesses to each levels of memory hierarchy. 
%The edge component, i.e., the inter-device communication overhead is calculated based on a simple throughput model. 

\vspace{-0.2cm}
\subsection{Memory Hierarchy Modeling} 
%re-streaming coupled with tiling
%As pointed out before, the success of roofline analysis very much depends on the accurate estimation of the number of accesses to each level of memory hierarchy. 
The number of accesses to each level of memory hierarchy is a function of the underlying hardware (memory capacity at each level) and the algorithmic implementation (loop ordering and tiling strategies). 

%For any given hardware configuration, at each level of memory hierarchy, 
For any given input configuration, we explore $N^L$ random tiling strategies which meet the memory capacity requirement at each level. 
$N$ is the number of tiling strategies at each level and $L$ is the number of levels of memory hierarchy. 
Empirically, we found that  for $L=3$, $N \approx 20$ results in a reasonably accurate estimation. 

For a given tiling strategy, it is easy to find the number of times each tile needs to be re-streamed from the next level of memory hierarchy. 
We start from the lowest level (main memory) and walk upward to estimate the number of accesses.
%(refer to Alg.~\ref{alg:}). .
The number of memory accesses at each level is dictated by the tiling strategy at current level and the higher level. 
For the highest level, the number of accesses is determined by the dataflow strategy exploited at MCU units. 
%Note here that $T_{(i)}$ is itself a set that captures tiling along different dimensions. For example, in the case of GEMM, $T_{(i)}$ would be a triplet with one value for $t_x$, $t_y$ and $t_z$, which captures tiling along the input, output and inner dimensions. 

%%%%%%%%
\subsection{DataFlow Model} 
%The number of accesses to each level of memory hierarchy, except for level 0 (i.e. register files) is dictated by the tiling parameters at the current and higher levels. 
%For register files, the number of accesses is dictated by the architecture of the underlying execution engine and the dataflow strategy that it employs. 
The number of accesses to the highest level of memory hierarchy (i.e. register file) will be determined by the number of instructions executed in the execution engine and the dataflow strategy governing mapping and communication between those engines (e.g. weight stationary, activation stationary and output stationary~\cite{eyeriss,timeloop}). 
The execution engine structure dictates how many times a piece of data could be reused internally before accessing the register file. We refer to this number as \textit{reuse factor (K)}.
%For example, for a fused multiply-add (FMA), each data element will be reused only once. 
In a 2-D systolic array with size $N_x$ and $N_y$, and an input GEMM with size $T0_x$, $T0_y$ and $T0_z$ at $L_0$, each data element could be reused $T0_x/N_x$ or $T0_y/N_y$ or $T0_z/N_z$ times, depending on which matrix is stationary. 
%We model three different dataflow strategies, weight stationary ($WS$), activation stationary ($AS$) and output stationary ($OS$) and allow users to pick their desired strategy from one of the $\{WS$, $AS$, $OS$, best$\}$ choices. Here, the \textit{best} will selects the best strategy among the three choices.
Given the reuse factor, we estimate the number of accesses to register files as follows:

\vspace{-0.2cm}
\small
\begin{equation}
\displaystyle
     \#RegAccess = \#Flops \times \frac{N_x.N_y + K. N_x + K. N_y}{2.K.N_x. N_y}
\end{equation}
\normalsize
%%%%%%%%%%
\vspace{-0.5cm}
\subsection{Inter/Intra-Package Communication Modeling}
%The inter/intra-package communication timing is calculated based on a simple throughput analysis:
%We use throughput analysis to calculate the inter/intra-package communication timing:
%So far we discussed how performance prediction engine predicts the time for each node within the compute graph. Here, we look deeply into how to predict timing for each edge.
%Once the transformed compute graph is mapped down to the system graph, we calculate the number of bytes that needs to be transferred through each edge ($D_e$). On the other hand, 
As discussed in Section~\ref{sec:mapping}, compute graph to system graph mapping captures logical edge to physical link mapping.  
The effective bandwidth for each link is downrated by the number of logical edges sharing the link.
%Based on how many logical edges share a physical link, we find the portion of the bandwidth of the physical link that is allocated for every edge sharing that link. 
%using equation~\ref{eqn:bandwidth_derating_factor}.
%For edges which are allocated multi-hop paths, 
%The physical link with minimum effective bandwidth 
%is the bottleneck bandwidth which we use for time estimation for all edges.
%allocation becomes the bottleneck. 
%We use the bottleneck bandwidth for time estimation for all edges.
%For edges which are allocated multi-hop paths, the physical link with the minimum bandwidth allocation becomes the bottleneck. 
%We then use the bandwidth allocated in the bottleneck link (equation~\ref{eqn:bottleneck_link}) to estimate the total time ($T_e$) taken to transfer the tensors from one device node to the other using equation~\ref{eqn:transfer_time}.

%\begin{equation}
%    \label{eqn:bandwidth_derating_factor}
%    B^l_e = \frac{B^l}{\sum_{e \in E} X^l_e} ~~~~\forall l\in L
%\end{equation}
%\begin{equation}
%    \label{eqn:bottleneck_link}
%    \displaystyle{B_e = \min_{l}(B^l_e)}
%\end{equation}
%\begin{equation}
%    \label{eqn:transfer_time}
%    T_e = \frac{D_e}{B_e}
%\end{equation}
%where $B^l$ is the total bandwidth of link $l$, $X^l_e$ is 1 if edge $e$ is assigned to link $l$, otherwise 0, $B^l_e$ is the bandwidth allocated in link $l$ to edge $e$, $B_e$ is the bottleneck bandwidth for edge $e$, $D_e$ is the total amount of data (tensor size) sent along edge $e$ from one device node to the other, and $T_e$ is the total transfer time.

\begin{figure*}[!htb]
\minipage{0.3\textwidth}
  \centerline{\includegraphics[width=0.88\linewidth]{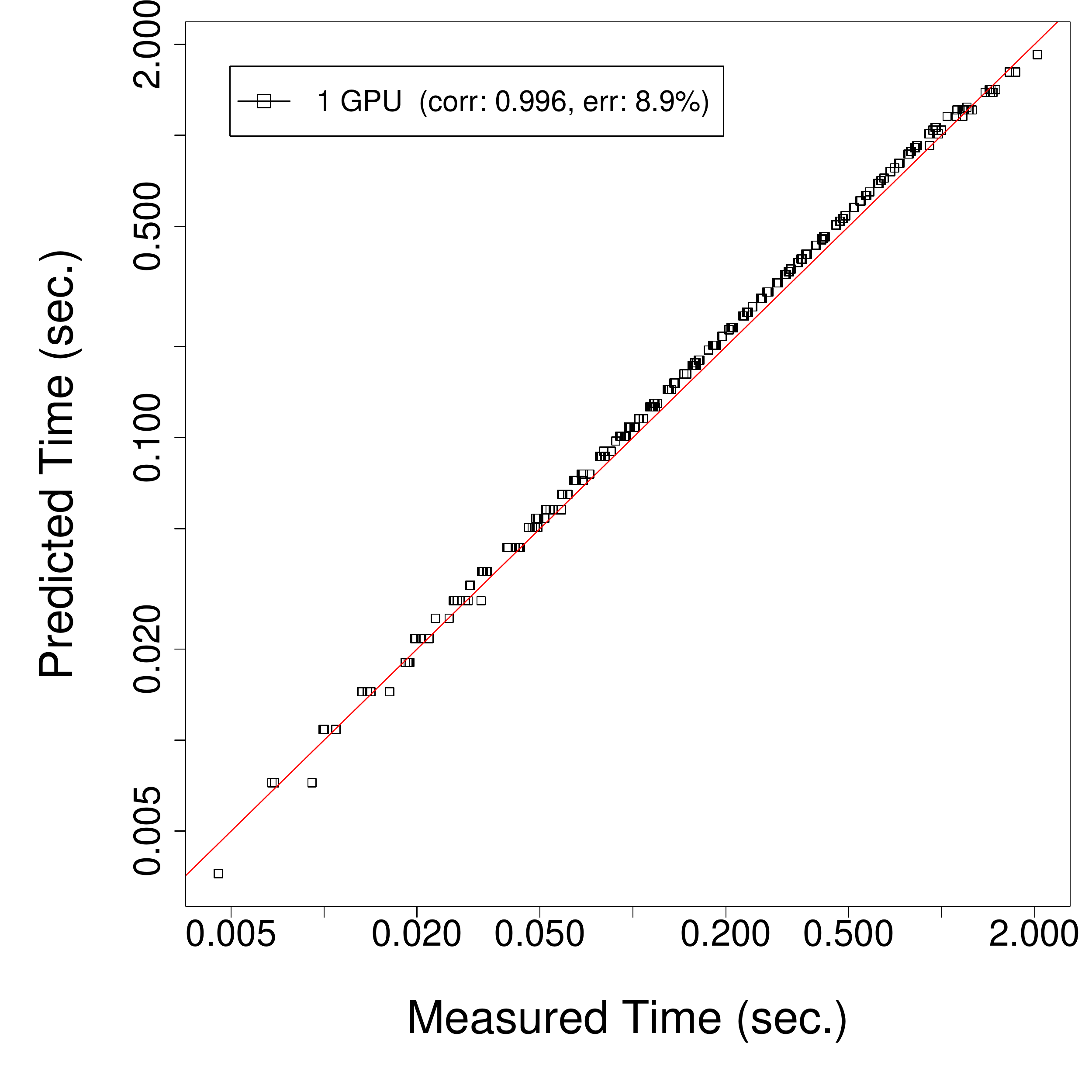}}
  \vspace{-0.3cm}
  \caption{GEMM Validation on P4.}
  \label{fig:gemm_p4}
\endminipage\hfill
\minipage{0.3\textwidth}
  \centerline{\includegraphics[width=0.88\linewidth]{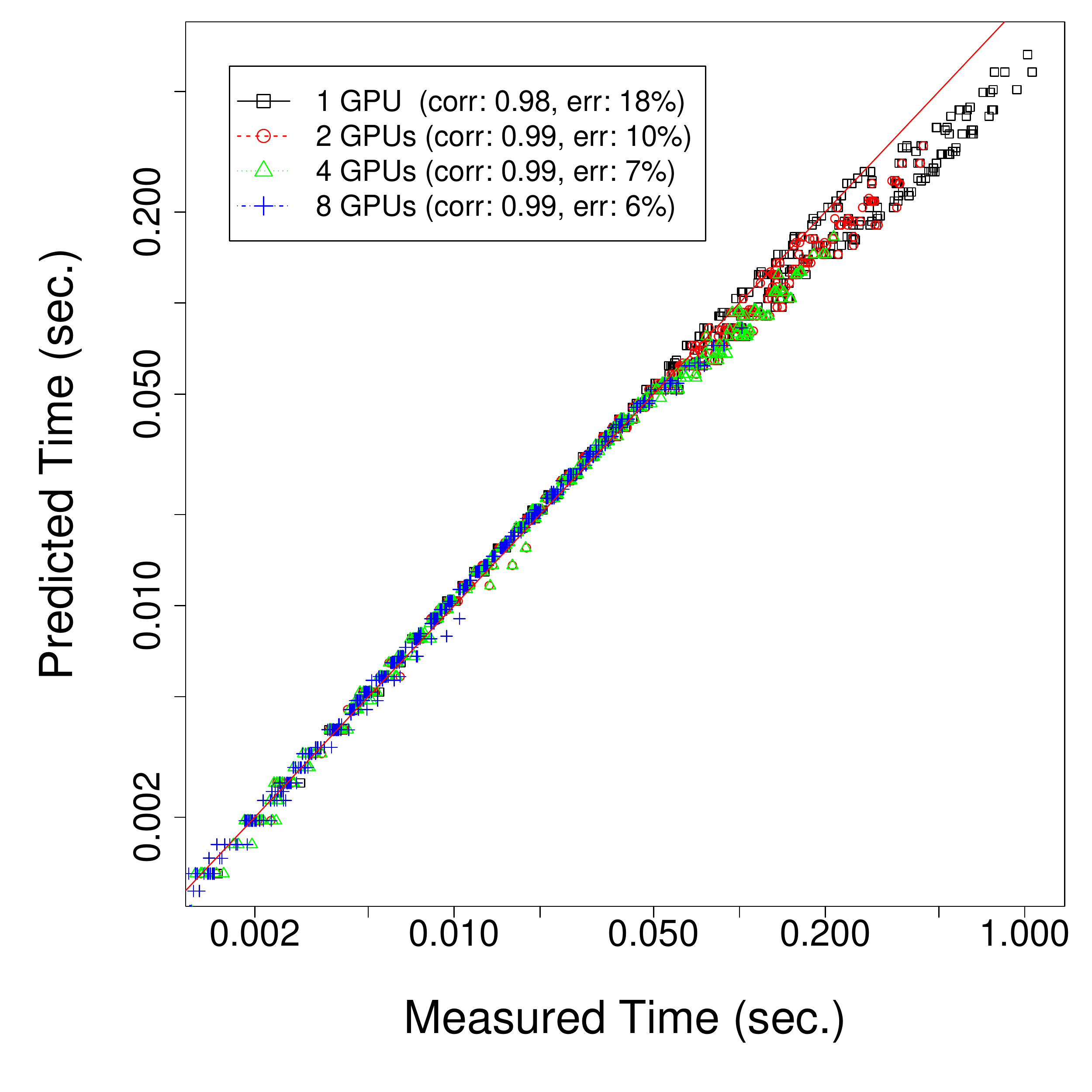}}
  \vspace{-0.3cm}
  \caption{GEMM Validation on DGX.}
  \label{fig:gemm_val}
\endminipage\hfill
\minipage{0.3\textwidth}
  \centerline{\includegraphics[width=0.88\linewidth]{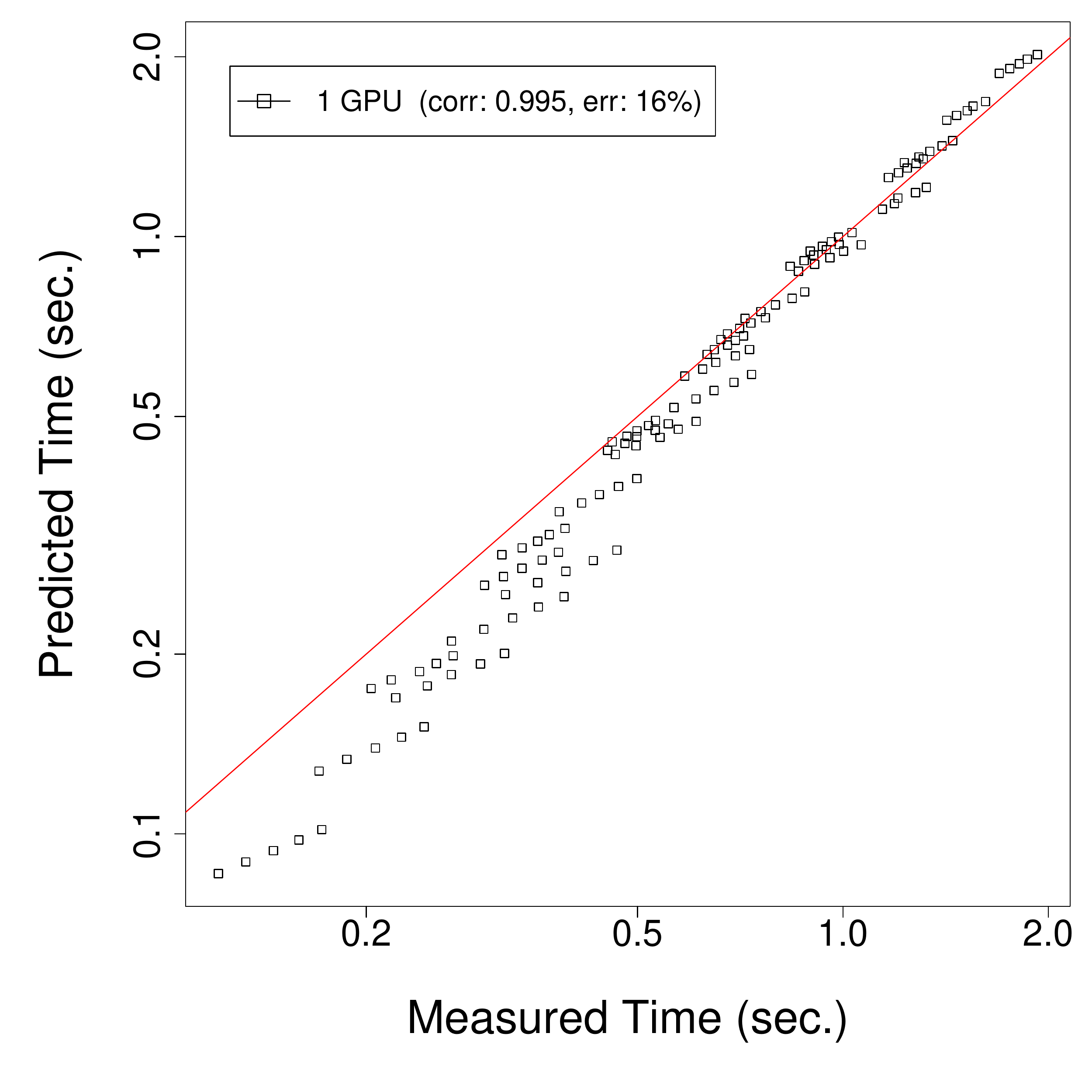}}
  \vspace{-0.3cm}
  \caption{LM Validation on V100.}
  \label{fig:lm_val}
\endminipage
\vspace{-0.1cm}
\end{figure*}
%%%%%%%%%
\vspace{-0.1cm}
\subsection{End-to-End Time Estimation} 
We use an event-driven simulation to estimate end-to-end timing. Event-driven simulation is basically a resource-constrained critical path analysis. Since multiple compute nodes can map into the same hardware node, event-driven simulation is necessary to avoid resource conflicts and respect resource scheduling constraints (e.g. not more than $k$ kernels can run in parallel on each hardware node).

We apply event-driven simulation at the original compute graph where the only parallelism to account for is pipeline parallelism: data parallelism and kernel parallelism would essentially create replicas of the original graph (where the kernel size and/or data size would be different for each node). 
Given that all replicas by definition are hosted on separate hardware nodes, they can all start and stop at the same time (assuming a homogeneous distribution of data along model replicas and homogeneous distribution of sub-kernels across data replicas) and their timing is deterministic. Hence, there is no need for event-driven simulation at the super-graph granularity. 

Figure~\ref{fig:transformation} explain an example of an end-to-end time estimation of a backward pass for a simple 3-layer feed-forward neural network, with 2-level pipeline parallelism ($p2$), 3-level data parallelism ($d3$), and 8-level kernel parallelism ($\texttt{R4-C2}$). %We start with kernel parallelism and then data parallelism to resolve the compute time for each node in the graph (this can be applied in any order). Once time for all nodes in the original compute graph is resolved, we use event-driven simulator to account for pipeline parallelism and resource scheduling constraints.
\section{Design Space Exploration Engine}\label{sec:dse}
%\textbf{Parameters} Input parameters, $W$ are grouped into three different categories: area ($A$), power ($P$) and perimeter ($R$). Parameters within each group capture the percentage of the overall area, power or perimeter allocated to each resource.
%\begin{equation}
%    W & = \{A, P, R\}
%\end{equation}
We denote the set of hardware \textbf{\textit{parameters}} to explore as $W = \{\{A_i\}_0^{H-1}, \{P_i\}_0^{H-1}, \{R_i\}_0^{H-1}\}$, 
where $H$ is the number of micro-architectural components in the hardware accelerator node, and $A_i$, $P_i$ and $R_i$ capture the percentage of the overall area, power and perimeter allocated to each component, respectively. 

Our \textbf{objective} is to find the optimal $W^*$ that minimizes the total run time, $f(W)$,
such that $\sum_{i=0}^{H-1} A_i \leq 1$, $\sum_{i=0}^{H-1} P_i \leq 1$, and $\sum_{i=0}^{H-1} R_i \leq 1$.
The objective function $f$ does not have a closed form, but we can calculate  it by  querying the performance model (CrossFlow).
This problem is an example of a \textit{constrained black-box continuous} optimization.
Since the objective function evaluation (i.e. querying CrossFlow) is considerably cheap (milliseconds), 
we use a variation of projected gradient descent (GD) optimization to solve for $W^*$ (see~\ref{eqn:gd_search}).
Empirically, we found that GD with exponential averaging in the parameter space (rather than gradients) works the best for our problem. %, which we refer to as value-momentum.

\small
\begin{equation}
\vspace{-0.5cm}
\begin{split}
    W_t & = W_{t-1} - \eta \mathrm{g_t} \qquad     \hat{W_t} = \frac{W_{t}}{||W_{t}||} \\
    M_t & = \beta M_{t-1} + (1 - \beta) \hat{W_t} \\ 
    W_t & = \textrm{Project}(M_t) \quad \textrm{onto} \quad C_A, C_P, C_R
\end{split}
\end{equation}
\label{eqn:gd_search}
\normalsize

Where $W_t$ and $\mathrm{g_t}$ are the input parameters and gradients at time step $t$, $\eta$ is the learning rate and $\beta$ is the discounting factor. 
%The intuition behind using momentum over values rather than gradient is to ...
We repeat the update steps shown above until convergence or the maximum number of steps ($T$), whichever conditions happens earlier. 
The final result is very sensitive to initialization. We repeat the steps above from $S$ different starting points and return the best result.
Empirically, we found that $T=100$ and $S=10$ are sufficient to find a near optimal solution.  

%\input{figs/validation}
%\input{figs/gemm_p4}
%\input{figs/gemm_val}
%\input{figs/lm_val}
%\input{figs/case}
%\vspace{-0.6cm}
\section{Validation}\label{sec:validation}
We validate our performance prediction model against execution time measured on real systems (Nvidia P4 with 1 GPU and an NVIDIA DGX-1 system with 8 V100 GPU cards), running distributed GEMM as well as large-scale language models. %unlike prior work that validate against a simulator, like Maestro???~\cite{})
%\subsection{Methodologies}
%\textbf{Applications} We look into important kernels like distributed GEMM, and important end-to-end applications like language model (RNN-based) for validation.
%For validation, we look into important \textbf{applications and kernels}, including distributed GEMM and large-scale language model (RNN-based).
In particular, we study (2-layer LSTM) language models (LM) for validation and case study as it is deemed to be one of the most challenging applications to scale~\cite{hestness2019beyond}, 
%Moreover, it is an essential component to many important industry products, including search and speech recognition and
and is very costly to train~\cite{dlcost}. All applications are implemented in Tensorflow 2.0.
%and has tight time-constraints; 
%it needs to be retrained very frequently and is very costly to train. All the application are implemented in Tensorflow 2.0.
%
%\textbf{Measured Time} 
%We collect the real execution times (\textbf{true labels}) by running the applications on two different system setup: 1. an Nvidia P4 card and 2. an NVIDIA DGX-1 system with 8 V100 GPU cards. 
%We report the average execution time of 100 runs. 
%For 1 GPU runs we use Nvidia profile (nvprof) to collect timing of GEMM kernels. 
%For more than 1 GPU and LM applications, we use Tensorflow 2.0 time measurement hooks to measure the training time for each epoch. 
%This number would include all the software stack latency.
%
%\textbf{Predicted Time} 
We use CrossFlow to predict the runtime, %of the applications we run on the hardware. 
which can take anywhere from milliseconds to 20 seconds. %depending on the kernel dimensions.
%We also adjust the maximum utilization parameters for each resource through micro-benchmark based measurements. 
%Using Nvidia system profiler tool and micro-benchmarks, we found that on V100 cards, cores, memory, caches and network operates in average at 85\%, 100\%, 65\% and 95\% of peak utilization, respectively. While the utilization number might change slightly from kernel to kernel or even different configurations of the same kernel, we found that variations in sustained throughput is negligible (within 2-5\%).

%\textbf{Validation Space} 
For GEMM validation, we look at a space of more than 2000 GEMM kernels of different shapes and parallelism strategies, where input (\texttt{m}), output (\texttt{n}) and inner dimensions (\texttt{k}) varying from 4K to 32K in steps of 4K, and parallelized across 1, 2, 4, or 8 GPUs, using both Row-Column and Column-Row distributed parallelism strategies. 
For LM validation, we look into a space of 125 configurations, where \texttt{Batch Size}, \texttt{Hidden Dimension} and \texttt{Vocab Size} varying from 2K to 6K in steps of 1K. 
%The range of the values to sweep through is more limited for LM due to its large memory capacity requirement and capacity limitation for measurement on real system.
%
%Note here that LM validation space for an end-2-end application is smaller than GEMM's due to large memory requirements for softmax and embedding layers during training on real hardware, and memory capacity limitations for measurement on the V100 GPUs in a DGX-1 system.
%
%\textbf{Prediction Accuracy Metric} 
We report the correlation (corr), and also the mean relative error (err) to quantify the quality of our predictions. %We refer to this value simply as error in the discussion follows. 

%\textbf{Results}
%One of the major components of any deep learning application is GEMM kernel.
%We study the performance prediction of GEMM kernel in the context of single and distributed GEMM implementation. In the distributed GEMM cases, we implemented two parallelization strategies: (1) Row-Column, and (2) Column-Row~\cite{}.
Figure~\ref{fig:gemm_p4} shows the validation results on Nvidia P4 GPU card. On the X-axis, we show the measured time (in log-scale), and on the Y-axis, we show the predicted time (in log-scale).
As shown, predictions and measurements are highly correlated (0.996) and the error is small (8.9\%).
Figure~\ref{fig:gemm_val} shows that CrossFlow  predictions  on a DGX-1 system across 1, 2, 4 and 8 V100 GPU cards are well correlated (0.98-0.99) and have low error (6\%-18\%).
%As shown, we can predict the run-time with 10-14\% average error across different GPU cards and different distributed training settings. 
%The error is mostly dominant in 
%\textbf{End-2-End Application Validation}
Figure~\ref{fig:lm_val} shows the performance of LM on V100 GPU card. Similarly, we can predict performance with high correlation (0.996), and low error (16\%).
A constant pattern visible across all results is the performance prediction deviation from measurement on real hardware for small kernels. This is  expected as Tensorflow 2.0 time measurement hooks include all the software stack latency; while this overhead is negligible for large kernels, it accounts for a large portion of total run-time if the kernel is very small.
This indicates the tool outcome would be more reliable for large kernels and large models.

%\input{figs/case}
%\vspace{-0.2cm}
\section{Case Studies}\label{sec:case-study}
%Here we will show how we use CrossFlow to project the impact of logic, network and memory technology scaling on the overall performance. 
%We also show the impact of co-optimizing parallelism strategy and the architecture at different technology nodes.
%\name is a very expressive model that not only can be used to predict performance but also to predict hardware counters (number of accesses to different levels of memory hierarchy for different operations, number of bytes transferred for each operation under each parallelism strategy, etc.) which we use to understand the observed pattern.  

%\subsection{Methodology}
DeepFlow is a pathfinding framework with studies and use cases spanning semiconductor technology development, micro-architecture, neural network models, and algorithmic parallelization techniques. In this section, we give few example case studies  for a large-scale language model 
(hidden dim: 16K, global batch size: 16K, vocab size: 800K, number of layers: 2, sequence length: 20) distributed across 512 hardware nodes. For future technology exploration, we study 7 consecutive \textbf{logic} technology nodes (from 12nm (N12) to 1nm (N1). Based on the recent scaling trends for logic technologies~\cite{stillmaker,Wikichip_technode}, we assume area and power scale by 1.8$\times$ and 1.3$\times$ from one node to the next for iso-performance), 4 different \textbf{memory} technologies (HBM2 (1 TB/s), HBM2e (2 TB/s), HBM3 (projected 2.6 TB/s~\cite{HBM3}), and HBM4 (projected 3.3 TB/s)) and 3 different \textbf{network} technologies (Infiniband-NDR-x8 (100 GB/s), XDR-x8 (200GB/s) and GDR-x8 (3.3 TB/s)). The caveat to these results (as with any pathfinding study with DeepFlow) is that if the system architecture or dataflow or neural network is radically different (e.g., this study assumes that same node is homogeneously replicated within the package),  the conclusions may change. 
 
%. Similarly, for memory and network technologies, we assumm that the architecture and signalling technologies will improve signalling bandwidth and the current and proposed scaling trends in each domain will hold in the future.

\subsection{Impact of Technology Scaling}
\begin{figure}[h]
  
\centerline{\includegraphics[width=0.9\linewidth]{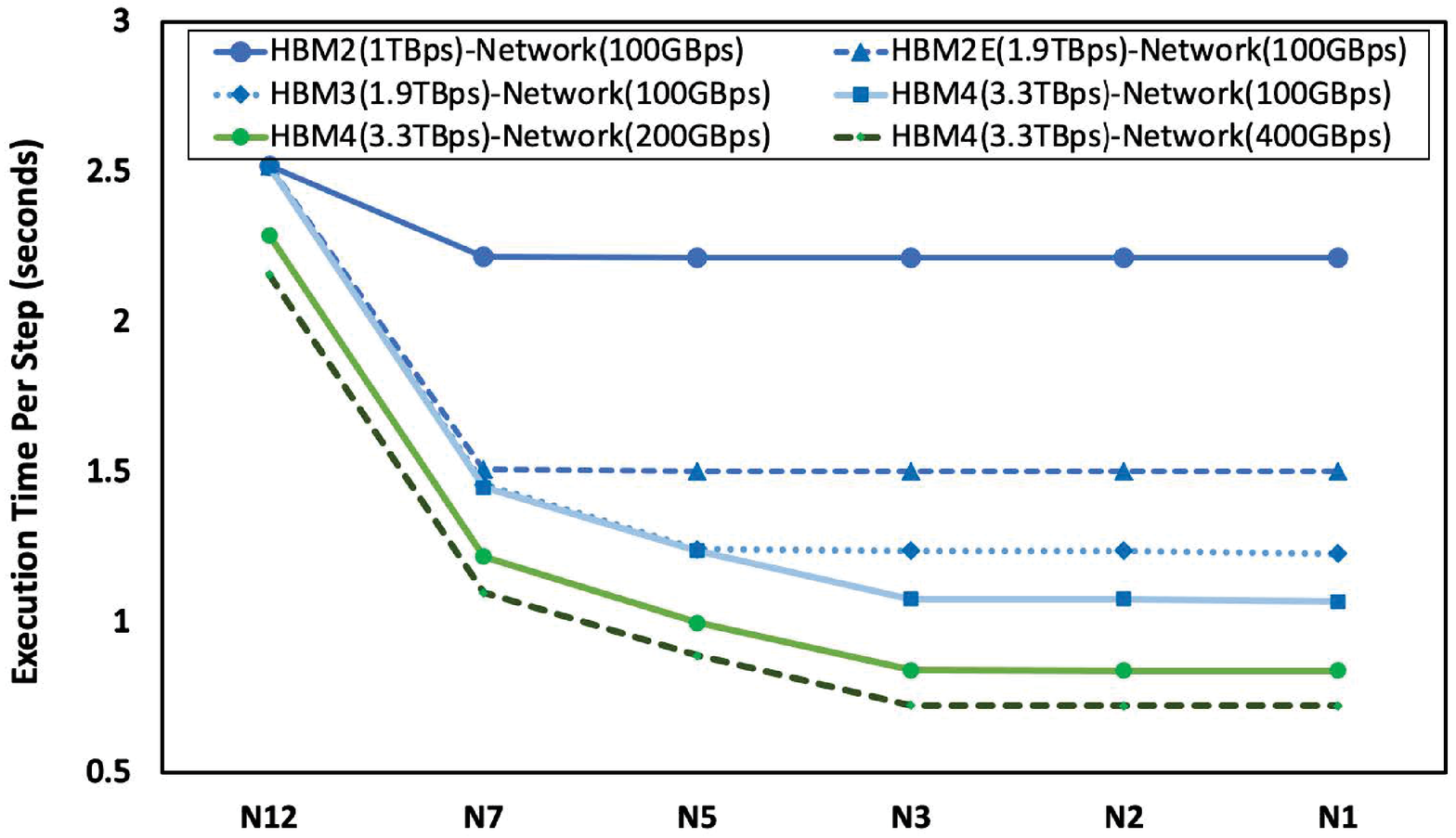}}
  \caption{\textbf{Technology Scaling:} scaling logic, memory and network technology}
  \label{fig:case1}
  \end{figure}
%\textcolor{blue}{Here we asked the question, scaling which technology would provide the maximum end-to-end performance benefit?} 
The first question we seek to answer is where the performance bottlenecks are across the stack and which technology could provide the maximum end-to-end performance benefit? Semiconductor technology development decisions are increasingly driven by machine learning as the workload. Many of these decisions trigger large, multi-year investments. Figure~\ref{fig:case1} shows the impact of scaling logic, memory, and network technology for a large-scale language model using data-parallelism.
For these experiments, we assume that power/node = 300W and area/chip = 850 $mm^2$. %, and the fraction of power and area devoted to each u-architectural component is fixed. %Also, we fix the parallelization strategy at data parallelism.

%As shown, on the X-axis, we improve logic technology nodes from 12nm to 20\AA, and on the Y-axis we report execution time. The top four lines correspond to four recent and upcoming memory technologies, and the bottom three lines correspond to new networking technologies that improve bandwidth. Within each line, we observe a saturating trend, indicating that logic scaling alone is not a sustainable approach: 
Logic scaling improves compute throughput, and also caching capacity and bandwidth, but only to a smaller extent. Going from N12 to N7, we observe a jump in performance irrespective of memory technology. This is because at N12, the performance of a significant number of kernels are L2 bandwidth bound. At N7, the L2 bandwidth and capacity improve enough for HBM bandwidth to become the new bottleneck. Therefore, with improvement in HBM bandwidth, the balance can shift back again to caches and saturation point can be further improved with logic scaling, hence saturation point shifts further to the right. This trend continues up to N3. Beyond N3, even at very high memory bandwidth (3.3 TB/s) and network bandwidth (400 GB/s) performance stays unchanged as cache capacity and bandwidth are the main bottlenecks. Since the on-chip network connecting MCUs to cache and the cache controller overhead scale along with number of cache banks and the number of MCUs (which scale at  $\sim1.8\times$ per technology node), the cache capacity as well as bandwidth increase only marginally at N2 and N1. These trends are well inline with commercial examples from NVIDIA and AMD, where jump to N7 node provided large performance benefits and then, multiple high-end SKUs of the GPUs with higher bandwidth HBM memories have been released for further performance improvements.

\textbf{Network technology} scaling is another big factor that determines overall end-to-end performance of a distributed deep learning system. As logic and memory technologies scale alongside the size of the models, more inter-node bandwidth is needed to accelerate the inter-node communication collectives. Our analysis (Figure~\ref{fig:case1}) shows that beyond N3, scaling networking technology will provide much larger performance gains as opposed to logic scaling. This trend also aligns with the recent efforts in the industry to push high bandwidth and low energy networking technologies and architectures for inter-node and intra-node communication, targeted towards deep learning systems~\cite{google-optical-1, google-optical-2, NVLINK}.

%but logic scaling cannot sufficiently improve cache capacity/bandwidth:
%
%
%However, we see that beyond N3, even 3.3 TB/s bandwidth wouldn't be sufficient to provide further performance improvement. 
%
%This lead us to find the cause behind why would increased cache capacity from advanced nodes don't provide performance benefits?
%This is because the overhead of the on-chip network connecting the MCUs to the cache %, and the cache controller circuitry overhead 
%scales up with the number of MCU cores and the number of banks. 
%Moving from one node to the next will increase the number of MCUs on chip by $\sim1.8\times$.
%As a result, the cache capacity as well as bandwidth increases only marginally.
%and as a result bottlenecks performance under fixed chip area and power budgets.

%Our results also suggest that as logic technology improves, the bottleneck shifts from compute to memory and as memory improves, the bottleneck shifts from memory to network. But beyond Infiband-GDR, we see that the improvement from network would be marginal. This is because the communication portion becomes a small fraction of the end-to-end time.
%Understanding where the bottlenecks are is of our paramount importance as it directly influence our strategy for scaling. CrossFlow enables such bottleneck analysis from technology-levels all the way to algorithmic levels.
%it is important to understand if %we are compute-bound, memory-bound or network-bound, as it demands different scaling solutions.

%\vspace{-0.1cm}
\subsection{Co-optimizing Technology, Parallelism Strategy and Hardware Architecture Design}
\begin{figure}[h]
    \centerline{\includegraphics[width=0.9\linewidth]{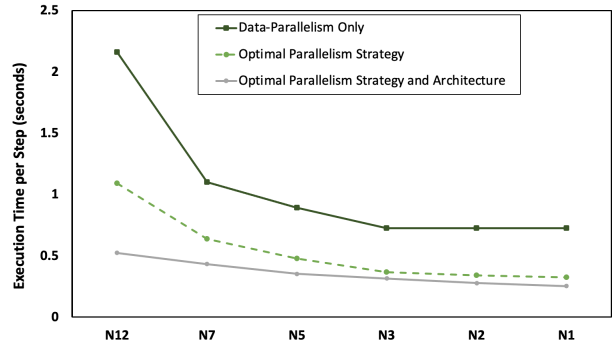}}
  \caption{Co-optimizing parallelism strategy and hardware architecture design.} 
  \label{fig:case2}
  
\end{figure}

% We argue that technology and architecture exploration should be co-optimized along with parallelism strategy.
Figure~\ref{fig:case2} shows the importance of co-optimizing technology with parallelism and hardware design in an incremental fashion.
%The top-line is the baseline with no architecture or parallelism strategy exploration (same as the green dashed line from Figure~\ref{fig:case1}). 
%
%The next line shows model's performance with the optimal parallelism strategy at each technology point. 
%Finally, the bottom line shows model's performance when co-optimizing parallelism strategy and hardware architecture together.
%
%As technology nodes get more advanced and compute throughput per node improves (with cache bandwidth and capacity to a smaller extent), the best parallelism strategy would also change. For example, at N12, the best parallelism strategy is $\texttt{R2-C4-D64-P1}$, while at N1, it is $\texttt{CR4-D128-P1}$. 
%
%Finally the last line shows model's performance when co-optimizing technology,  parallelism strategy and hardware architecture together.
%
As shown: 
(1) Parallelism strategy optimization alone can offer $\sim2\times$ performance improvement.
(2) Co-optimizing architecture and parallelism strategy offers meaningful benefits for mature technology (12nm and 7nm) nodes. But for more advanced technology nodes, only marginal benefits (20\%-30\%) can be gained on top of parallelism strategy optimization. 
(3) For current and near-future technology nodes, co-optimizing for model architecture can provide as much benefit as scaling technology nodes (by almost two generations). 

%\vspace{-0.15cm}
\subsection{Effect of Multi-Node Package}
\begin{figure}[t!]
   
\centerline{\includegraphics[width=0.8\linewidth]{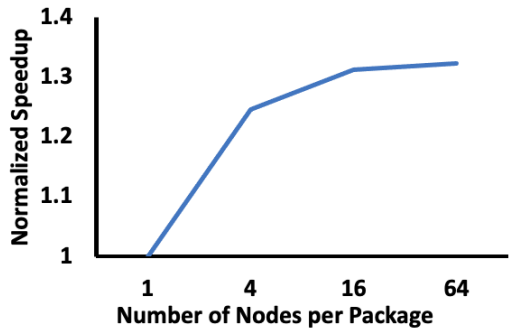}}
  \caption{Performance improvement from multi-node package} 
  \label{fig:multi-node_package}
%\caption{Best case execution time improvement from multi-node packages}
\end{figure}

Next, we evaluate the performance improvement that multi-node packaged systems (e.g., MCM-GPU~\cite{mcm-gpu}, waferscale-GPU~\cite{ws-gpu}, Tesla Dojo~\cite{tesla-dojo}) can provide in a distributed training setup (see Fig.~\ref{fig:multi-node_package}).
We assumed 2TB/s link bandwidth for the intra-package links and performed both parallelism and architecture search for each case.

%\begin{table}[]
%\centering
%\footnotesize
%\begin{tabular}{|c|c|}
%\hline
%\#Nodes in Package & Normalized Execution Time \\
%\hline
%1                  & 1                         \\
%\hline
%4                  & 0.802397143               \\
%\hline
%16                 & 0.761674856               \\
%\hline
%64                 & 0.755620015              \\
%\hline
%\end{tabular}
%\caption{Best case execution time improvement from multi-node packages}
%\end{table}
Couple of key takeaways from these experiments were: (1)  Increasing the number of nodes in a package improves overall performance by roughly 32\% at best. (2) Beyond 4-nodes per package, performance improvement is marginal. Since ultra-large packages or waferscale integration dramatically worsens cost, we believe that such technologies may not be worthy investments for scaling large language model training. These conclusions hold across multiple different batch sizes, hidden dimension sizes and intra-node link bandwidths.

%To conclude, we argue that building optimized architectures at relatively mature nodes (N12/N7) can provide as much benefits as building architectures at the more costly advanced nodes. On the other hand, support for different parallelization strategies and algorithms to guide parallelism choice need to be enabled so that ML practitioners can derive optimal performance out of a given hardware. 

%We use that as a baseline where no architecture exploration conducted.
%We then show how parallelism exploration 
%Figure~\ref{fig:case2} (the dark green line) is same as the green dashed line from Figure~\ref{fig:case1}. 

\section{Related Work}
%\vspace{-0.2cm}
Related work can be broadly categorized into (1) performance modeling frameworks for spatial architectures like TimeLoop and Maestro, (2) performance modeling frameworks for parallelism exploration such as FlexFlow, and (3) what-if analysis tools like DayDream and Habitat.

Similar to TimeLoop~\cite{timeloop} and Maestro~\cite{maestro}, we use an analytical model to estimate performance, however, the scope of DeepFlow is much broader. 
TimeLoop and Maestro model a single kernel runtime on the spatial architecture like systolic array or Eyeriss. Similarly, Mind Mapping~\cite{mindmapping} is a gradient based search tool that finds the best tiling and mapping strategy for a single compute unit and is built on top of Timeloop.
In this regard, all these prior work are similar to analytical models that goes into DeepFlow's MCU modeling. However, DeepFlow offers more than MCU modeling.
DeepFlow allows to capture not only the behavior of an MCU unit but also an entire GPU (through modeling of communication across MCU units through shared layers of memory hierarchy) as well as modeling a data center full of GPUs. Besides, prior work validates against simulators on micro-kernels. We validate our model against SOTA GPU hardware on real-world applications.
Furthermore, DeepFlow models an entire compute graph, composed of many kernels mapped and distributed across multiple GPU nodes, and allows the analysis of parallelism at this level, including pipeline, data and kernel parallelism. Moreover, DeepFlow provides four degrees of freedom to explore: model architecture, hardware architecture, technology configuration and parallelism strategy.

FlexFlow~\cite{flexflow} is an ML-based model for exploring the best parallelism strategy which relies on the runtime profiling tools to measure kernel timings on the target hardware. While it provides a very rich input for expressing different model architectures, it can only model existing hardware, hence not suitable for parallelism-architecture-technology co-design exploration. 

DayDream~\cite{zhu2020daydream} is a what-if analysis tool that enables researchers to evaluate the efficacy of different \textit{algorithmic} optimizations for an \textit{existing} hardware. However, it relies on fine-grain profiling tools to construct dependency graph, hence it lacks the ability
to predict individual kernel run-time on non-existing hardware and cannot be used for architecture or technology co-design space exploration.
Similarly, Habitat~\cite{geoffrey2021habitat} predicts deep learning workloads' run-time across different \textit{existing} GPUs, using a combination of wave scaling and MLP predictors. Wave-scaling can only model simple uarchitectural modification, and MLP predictors are u-architecture specific models that require collecting a large set of runtime data on the baseline and target hardware for model training, hence cannot be applied to non-existing hardware.

Astra-sim~\cite{astra-sim} is a simulator for hardware-software co-design of distributed deep learning systems. The focus of the paper is on detailed modelling of the inter-node network and they study the effects of network topologies and architecture choices. Astra-sim doesn't explore automated technology and architecture exploration and may not be suited for across the stack design space exploration because of the detailed and heavy-weight focus on network effects.

\vspace{-0cm}
\section{Conclusion}

%This paper is the first effort to explore the cross-stack impact of technology scaling, model scaling and architecture innovations from a holistic perspective, and at the same time considering real-world design constraints like area and power budget for deep learning training.
We proposed DeepFlow, a performance modeling framework that enables a cross-stack analysis for hardware-software-technology co-design at-scale. We envision DeepFlow to be used by \textit{ML practitioners} (to decide what hardware to use to maximize their utilization, or simply predict their hypothetical model architecture performance which might not be realizable in today's hardware for many reasons including capacity limitation), by \textit{system designers} (to decide what hardware accelerators they need to acquire or build from scratch to meet their application needs, what new technologies to invest in, etc.), and finally by \textit{ technology experts} (to guide future technology development by assessing its impact  all the way across the stack, at scale). Our future work plans to extend DeepFlow modeling to other applications beyond language models and GEMM kernels. 

%and, at the same time, considering .... Furthermore, we capture ... and present an end-to-end analysis for \textit{what} and \textit{how} hardware-software design and at-scale optimization can help improve system utilization. We share the key challenges and chart out important directions across all dimensions of AI---data, algorithms, systems, metrics, standards, and best experimentation practices.

%%%%%%%%% -- BIB STYLE AND FILE -- %%%%%%%%
%\clearpage
\bibliography{refs}
\bibliographystyle{ieeetr}
%\clearpage
%\appendix
%\section{A}
%
%\input{related}
%\input{appendix_hw_generator}

\end{document}